\documentclass[aip,jmp,amsmath,amssymb,preprint,onecolumn]{revtex4-1}

\usepackage[USenglish]{babel}
\usepackage{varioref}
\usepackage[T1]{fontenc}
\usepackage{ae}
\usepackage{fancyhdr}
\usepackage[latin1]{inputenc}
\usepackage{subfigure}
\usepackage{color}
\usepackage{appendix}

\usepackage{graphicx}
\usepackage{psfrag}

\usepackage{amsmath}
\usepackage{amssymb}
\usepackage{amsthm}
\usepackage{color}
\newcommand{\eqr}[1] {(\ref{eq:#1})}
\newcommand{\td}[2] {\frac{d #1}{d #2}}

\newcommand{\pd}[2] {\frac{\partial #1}{\partial #2}}

\newcommand{\md}[1] {\frac{D #1}{D t}}
\newcommand{\dmd}[1] {\frac{D #1}{D t^*}}

\renewcommand{\vec}[1]{{\bf#1}}
\newcommand{\n}{\noindent}

\newcommand{\DS}{\displaystyle}

\renewcommand{\div}{\vec{\nabla \cdot}}
\newcommand{\ddiv}{\vec \nabla^{\,*} \cdot}  
\newcommand{\dgrad}{\vec \nabla^{\,*}}  

\newcommand{\grad}{\vec \nabla }

\newcommand{\R}{\rho}
\newcommand{\T}{\theta}
\renewcommand{\u}{{\cal U}}

\newcommand\Rey{\mbox{\textit{Re}}}  
\newcommand\Pran{\mbox{\textit{Pr}}} 

\begin{document}

\preprint{}

\title[Mixing in Low-Mach number supercritical jets]{Turbulent mixing of a slightly supercritical 
Van der Waals fluid at Low-Mach number}

\author{F.~Battista}
\affiliation{Department of Mechanical and Aerospatial Engineering, Sapienza University, 
via Eudossiana 18, 00184 Rome, Italy}
\author{F.~Picano}
\affiliation{Department of Industrial Engineering, University of Padova, via Venezia 1, 35131, Padova, Italy}
\author{C.M.~Casciola}
\affiliation{Department of Mechanical and Aerospatial Engineering, Sapienza University, 
via Eudossiana 18, 00184 Rome, Italy}

\date{\today}

\begin{abstract}
Supercritical fluids near the critical point are characterized by liquid-like 
densities and gas-like transport properties. These features are purposely 
exploited in different contexts ranging from natural products 
extraction/fractionation to aerospace propulsion. 
Large part of studies concerns this last context, focusing on the dynamics 
of supercritical fluids at high Mach number where compressibility and
thermodynamics strictly interact.
Despite the widespread use also at low Mach number, the turbulent mixing 
properties of slightly supercritical fluids have still not investigated in detail in this regime.
This topic is addressed here by dealing with Direct Numerical Simulations
(DNS) of a coaxial jet of a slightly supercritical Van der Waals fluid. 
Since acoustic effects are irrelevant in the Low Mach number conditions found 
in many industrial applications, 
the numerical model is based on a suitable low-Mach number expansion of the 
governing equation. 
According to experimental observations, the weakly supercritical regime 
is characterized by the formation of finger-like structures-- the so-called 
ligaments --in the shear layers separating the two streams. 
The mechanism of ligament formation at vanishing Mach number 
is extracted from the simulations and a detailed statistical characterization is provided. 
Ligaments always form whenever a high density contrast occurs, independently 
of real or perfect gas behaviors. 
The difference between real and perfect gas conditions is found in the 
ligament small-scale structure. More intense density gradients and thinner 
interfaces characterize the near critical fluid in comparison 
with the smoother behavior of the perfect gas. A phenomenological interpretation is here
provided on the basis of the real gas thermodynamics properties.

\end{abstract}



\maketitle
\section{Introduction}
\label{sec:crit_jet}
A supercritical fluid is a phase of matter with no sharp transition between high, liquid-like density states and low, gas-like density states that exists  at pressures and  temperatures higher than those of the critical point. It consists  of a unique hybrid state intermediate between liquid and gas where no surface tension  acts at density interfaces. In certain regions of the phase diagram, supercritical fluids exhibit liquid-like density and gas-like transport 
properties  that diverge approaching the critical point.
These peculiar features make supercritical fluids attractive in several 
industrial and technological applications from space propulsions to chemical 
extraction processes~\cite{palmer1995applications,perrut2000supercritical,fages2004particle,brunner2010applications}.

The frequent use of supercritical fluids in aerospace propulsion devices, 
as in liquid rocket engines, motivated a substantial part of the studies in the
literature. Indeed, numerous experimental investigations on supercritical 
fluids~\cite{chetalcoy,segpol,royjolseg,MayTelBraSChHus,mayschschsch} have been 
addressing turbulent mixing properties, like Nitrogen/Heptane systems in experiments 
devoted to the basic understanding of mixing or Hydrogen/Oxygen systems for applications to combustion. 
Interesting reviews on these issues are  Refs~\onlinecite{bellan2000supercritical} and \onlinecite{zonyan} which  provide 
the state-of-the-art up to 2000.
In recent years numerical simulations also addressed supercritical mixing by investigating temporal mixing layers~\cite{MilHarBel,okobel,okobel1} and turbulent jet flows~\cite{oef}. 
Aiming at aerospace propulsion applications, all these studies considered supercritical flows with 
moderately high Mach numbers where acoustic effects and pressure fluctuations are crucial~\cite{MilHarBel}.

On the other hand many technological applications often employ fluids at low speed.
In industrial applications, supercritical fluids  are frequently used in 
place of CFC for cooling~\cite{linghejon}, to sterilize biological 
materials~\cite{checinska2011sterilization}, for textiles cleaning, and 
for electronics components degreasing~\cite{weibel2003overview}. 
Supercritical fluids are also widely adopted for chemical extraction, 
e.g.\ to extract substances from foods~\cite{king1996supercritical,mendiola2013supercritical} in processes like decaffeination by $CO_2$~\cite{capuzzo2013supercritical}.
For these cases at vanishing Mach number, where acoustic effects  are negligible, much less attention has been paid to turbulent mixing between a faster and a slower stream of even a single-component supercritical fluid.
It turns out that, when dealing with supercritical flows, the  computational difficulties known to arise 
in strongly subsonic flows, notably the increasingly stiff behavior of the compressible Navier-Stokes 
equations at increasing sound speed~\cite{majset,vol}, still need to be addressed in detail.
A major aim of the present work is to develop a consistent treatment of a supercritical fluid at small Mach 
number capable of being implemented in efficient numerical simulations of  turbulent mixing.

In this context, the \emph{Low-Mach expansion}, developed by 
Majda \& Sethian~\cite{majset} for perfect-gas reacting flows,
is certainly a fundamental starting point. However their 
formulation needs to be substantially rearranged to allow the extension to real gas 
conditions. The original part of the relevant asymptotic expansion
is presented here by focusing on the special case of Van der Waals fluids. 
It is then a straightforward exercise 
adapting it to other cases of practical interest, like to the widespread 
Peng-Robinson equation of state~\cite{penrob}.

Beside introducing the low Mach number expansion, the primary intent of the paper is investigating the peculiar features of the turbulent mixing of a single-component Van der Waals fluid at vanishing Mach number and slightly supercritical pressure. The issue is addressed by focusing on a co-axial jet with two streams at different  temperatures and velocities. Data from Direct Numerical Simulations (DNS) of real and perfect gas are compared to highlight the effect induced by the real gas thermodynamics.   Similarly to the high Mach number case~\cite{royjolseg,MayTelBraSChHus},  the weakly supercritical regime is characterized by the formation of finger-like structures (ligaments)  in the shear layers separating the two streams.  We show that ligaments always form whenever a high density contrast occurs, independently  of real or perfect gas conditions. The ligament small scale structure is  found to distinguish the real from the perfect gas behavior with more intense density gradients and thinner  interfaces characterizing the near-critical fluid  in comparison 
with the smoother behavior of the perfect gas.

The paper is organized as follows. Section \S~\ref{sec:method} is dedicated 
to the generalized Low Mach number expansion for real-gas  flows. Section 
\S~\ref{sec:VanderWaals} deals with the specific aspects of the Van der Waals model.
In section \S~\ref{sec:variable_density_jet} the variable-density turbulent 
coaxial jet simulations are presented, discussing the relevant aspects of the solution algorithm. 
The detailed physics of the near-critical jet dynamics is illustrated in \S~\ref{sec:results} that 
provides the main physical results of the paper. Final comments and conclusions are eventually 
reported in \S~\ref{sec:conclusions}. 
A few Appendices are included at the end, to allow the illustration of certain technical aspects without 
interrupting the main stream of the discussion.

\section{The Low-Mach number formulation for a generic equation of state}
\label{sec:method}

In the original derivation of the low-Mach number approximation of the fully 
compressible, reacting Navier-Stokes equations, Majda \& Sethian~\cite{majset} 
employed the perfect gas 
equation of state to describe the thermodynamic behavior of the fluid.
The original procedure described  in that seminal paper heavily relies on the particularly 
simple and specific form of the equation of state. 
However, when dealing with 
near-critical conditions, the equation of state should be generalized to treat more general 
cases and allow the use of the Van der Waals model or other appropriate 
real gas descriptions. 

Since deriving the approximation under these more general 
conditions requires a different manipulation of the equations, the basic procedure is here 
briefly  outlined.

The fully compressible, single component Navier-Stokes equations read as
\begin{align}
\label{N-S}
&\DS{\pd{\R^*}{t^*}} + \ddiv (\R^* \vec u^*) =0 \, , \\
\notag
&&\\
&\DS{\pd{(\R^* \vec u^*)}{t^*}} + \ddiv (\R^* \vec u^*\otimes\vec u^*) =
\ddiv \vec{\Sigma}^* - \dgrad p^* + \vec f^* \, , \\
\notag
&&\\
\label{eq:int_ener_dim}
&\R^* \left(\pd{\u^*}{t^*}+\vec{u}^*\cdot \dgrad{\u^*}\right)=
-p^*\ddiv \vec{u}^* +\vec{\Sigma}^*:\dgrad\otimes\vec{u}^* - \ddiv \vec{q}^* \, ,\\
\notag
&\\
\label{eq:scal}
&\DS{\pd{(\R^* Y^*)}{t^*}} + \div (\R^* \vec u^* Y^*) =
\ddiv ({\cal D}^* \dgrad Y^*) \, ,\\
\notag&\\
\label{eq:dstate}
&p^*=p^*(\T^*,\R^*) \, , \\
\notag&\\
\label{eq:ustate}
&\u^*=\u^*(\T^*,\R^*) \, , 
\end{align}
where the asterisk denotes dimensional variables.  In the equations, $t^*$, $\R^*$, 
$p^*$, $\T^*$, $\vec u^*$ are time, density, pressure, temperature and 
velocity, respectively with $\vec \nabla^*$ the spatial gradient and $\otimes$ the tensor product.
$\vec{\Sigma}^*= \mu^* [(\dgrad\otimes \vec u^* +(\dgrad\otimes \vec u^*)^T) 
-\left({\mu_B^*}/{\mu^*}- {2}/{3} \right) (\ddiv \vec u^*) \, \vec I]$
is the viscous stress tensor,   $\mu^*(\T^*)$  and $\mu^*_B(\T^*)$ are the 
temperature dependent dynamic and bulk viscosity, respectively,
and $\vec f^*=-\R^* g^* \vec e_z$ is the  gravitational force (with 
$\vec e_z$ the vertical unit vector). For simplicity, in eq.~\eqr{int_ener_dim} 
the heat flux $\vec{q}^*$ is assumed to follow the Fourier law, 
$\vec{q}^* = - k^*(\theta^*) \, \dgrad \theta^*$ as appropriate for single 
component fluids. $\u^*$ is the internal energy per unit mass, obeying an 
equation of state in terms of temperature and density, eq.~\eqr{ustate}.
Equation~\eqr{scal} is the convection-diffusion equation for a generic passive 
scalar, like a tracer that is mixed by the flow. Finally, eq.~\eqr{dstate} is 
a generic equation of state expressing the pressure in terms of density $\R^*$ 
and temperature $\T^*$. The system can be easily extended to deal with 
multi-component reactive mixtures by including additional convection-diffusion-reaction 
equations for each species and by considering a more complete transport model 
for heat and mass fluxes, see e.g. Ref~\onlinecite{law}. The extension of the Low Mach 
number expansion to be introduced below to the more general multi-component, 
reactive case is straightforward and will not be discussed further in the following.

The Low Mach number expansion is better performed starting from the dimensionless 
system. After selecting  a characteristic length $\ell^*_R$, a speed $|\vec{u}^*_R|$, a 
pressure $p^*_R$ and a density $\R^*_R$,  the other reference quantities follows as 
\begin{equation}
\label{eq:reference}
t^*_R=\frac{l^*_R}{|\vec{u}^*_R|}, \quad 
\u^*_R=\frac{p^*_R}{\R^*_R}, \quad
\T^*_R= \T^*_R\left(p^*_R, \rho^*_R \right)
\end{equation}
where the characteristic temperature $\T^*_R$ is expressed as a function of reference 
pressure and density through the pressure equation of state \eqref{eq:dstate}, see also 
\S~\ref{sec:VanderWaals}. The dimensionless system reads
\begin{align}
\label{massa_n_s}
&\DS{\pd{\R}{t}} + \div (\R \vec u) = 0  \\
\notag
&\\
\label{qta_moto_n_s}
&\DS{\pd{(\R \vec u)}{t}} + \div (\R \vec u\otimes \vec u) =
\frac{1}{\Rey}\div \vec{\Sigma} - \frac{1}{\gamma_R \widehat{Ma}^2}\grad p - \frac{1}{Fr^2}\R 
\vec e_z  \\
\notag
&\\
\label{energia_n_s}
&\R \left(\pd{\u}{t}+\vec{u}\cdot\grad{\u}\right)=
-p\div \vec{u} +\frac{\gamma_R \widehat{Ma}^2}{\Rey}\vec{\Sigma}:\grad\otimes\vec{u} + 
{\frac{c_{pR}^*}{{\cal Z R}_m^*}}\frac{1}{\Rey\,\Pran}\div\left(k \grad\T\right) \\
\notag
&\\
\label{specie_n_s}
&\DS{\pd{(\R Y)}{t}} + \div (\R \vec{u} Y) =
\frac{1}{\Rey Sc} \div ({\cal D} \grad Y)  \\
\notag
&\\
\label{stato_n_s}
&p=p(\T,\R) \\
\notag \\
\label{statoU_n_s}
&\u = \u(\T,\R) \, ,
\end{align}
where the relevant dimensionless parameters are
\begin{align}
\label{eq:parameters}
\notag
\widehat{Ma}= \frac{\left|\vec{u}^*_R\right|}{\sqrt{{\gamma_R p^*_R}/{\R^*_R}}},\quad 
\Rey=\frac{\left|\vec{u}^*_R\right| \, \ell^*_R \, \R^*_R}{\mu^*_R} , \quad
\Pran&=\frac{{c_p}^*_R \mu^*_R}{k^*_R},\\
&\\
\notag
Fr=\frac{\left|\vec{u}^*_R\right|}{\sqrt{\ell^*_R g^*}},\quad 
Sc=\frac{\mu_R}{{\cal D}_R \rho^*_R},&
\end{align}
with ${c^*_p}_R= c^*_p(\theta_R^*,\rho_R^*)$ being 
$c^*_p = \partial \left[{\cal U}^*+ p^*/\rho^* \right]/\partial \theta^*_{|p^*}$ the 
constant-pressure specific heat coefficient, ${\cal R}_m^*$ is the gas 
constant, $Z=p_R^*/({\cal R}_m^* \T_R^* \R_R^*)$ is the compressibility factor,
and $k^*_R$ and  $\mu^*_R$ the thermal diffusion coefficient and  the
dynamic viscosity, respectively,  evaluated at the reference temperature $\T^*_R$.  
$\widehat{Ma}$ takes the role of the Mach number, even if the quantity in the 
denominator does not directly coincide with the sound speed at reference thermodynamic 
conditions, since in general,  for a real gas, $c_R^2 = \partial p^*/\partial \rho^*_{|S^*}
(\theta_R^*,\rho_R^*) \ne \gamma_R p_R^*/\rho_R^*$ with $\gamma_R$ the ratio of 
constant pressure to constant volume specific heat and $c_R$ the actual sound speed in 
the reference conditions.

The Low Mach number formulation, Ref~\onlinecite{majset}, amounts to introducing the 
asymptotic expansion 
\begin{equation}
\vec f(\vec x , t)= \vec f_0(\vec x , t)+ 
\vec f_2(\vec x , t) \widehat{Ma}^2+ O(\widehat{Ma}^4) \, .
\label{expansion}
\end{equation}  
for the generic variable into system~\eqref{massa_n_s}-\eqref{statoU_n_s}. After grouping 
together terms with the same power in the Mach number and requiring that the resulting 
equations should be identically satisfied for any, sufficiently small Mach number,  
the system of equations governing the different terms in expansion~\eqref{expansion} 
 is readily obtained.

The zero-th order contribution to the mass conservation equation is
\begin{equation}
\DS{\pd{\R_0}{t}} + \div ({\R}_0 {\vec u}_0) = 0 \ .
\label{massa_exp}
\end{equation}
The same procedure applied to the momentum conservation  provides  a first 
contribution formally diverging like $1/\widehat{Ma}^2$.
Removing this low Mach number divergence yields the equation
\begin{align}
\label{grad_pres}
\grad p_0 &= 0 \,\Rightarrow\, p_0=p_0(t) \, ,
\end{align}
that implies a spatially constant zero-th order (thermodynamic) pressure. A second 
contribution arises at zero-th order in the Mach number and is  given by
\begin{align}
\DS{\pd{({\R}_0 {\vec u}_0)}{t}} + \div ({\R}_0 {\vec u}_0 \otimes {\vec u}_0) &=
\frac{1}{\Rey}\div \vec \Sigma_0 - \grad p_2 + \frac{\rho_0}{Fr^2} \vec e_z\, ,
\label{eq:qta_moto}
\end{align}
where  $p_2$ is  the second order (hydrodynamic) pressure~\cite{majset}.
Analogously,  the zero-th order equation for the transported scalar follows as
\begin{align}
\label{specie_n_s_bis}
\DS{\pd{(\R_0 Y_0)}{t}} + \div (\R_0 {\vec u}_0 Y_0) &=
\frac{1}{\Rey Sc} \div ({\cal D} \grad Y_0) \ .
\end{align}

In order  to complete the asymptotic expansion for real gases it is worth recasting the energy equation
\eqref{energia_n_s} in terms of temperature exploiting the equation of state \eqref{statoU_n_s} (more details
are given in Appendix~\ref{app:II}),
\begin{align}
\R \frac{c_v^*}{{\cal R}_m^*} \md{\T}=- Z \T \left.\pd{p}{\T}\right|_\rho \,\div \vec{u}
 +Z \frac{\gamma_R \widehat{Ma}^2}{\Rey}\,\vec{\Sigma}:\grad\vec{u} + 
\frac{c_p^*}{{\cal R}_m^*} \frac{1}{\Rey\,\Pran}\div\left(k \grad\T\right) \, ,
\label{eq:ener_evol_V}
\end{align}
where $c_{v}^*=\left. \partial{\u^*}/\partial{\T^*}\right|_{v^*}$ is the constant-volume specific heat coefficient.
Exploiting the Low-Mach number expansion of the temperature equation,
the zero-order contribution follows as
\begin{align}
\R_0  \frac{c_{v0}^*}{{\cal R}_m^*}\md{\T_0}=
-Z \T_0 \left.\pd{p_0}{\T_0}\right|_{\R_0} \,\div \vec{u_0}
 + \frac{c_{p0}^*}{{\cal R}_m^*}\frac{1}{\Rey\,\Pran}\div\left(k_0 \grad\T_0\right)
\label{eq:ener_evol_VI}
\end{align}
where $c_{v0}^*=c_{v}^*\left(\T^*_R \T_0,\R^*_R \R_0\right)$ and 
$c_{p0}^*=c_{p}^*\left(\T^*_R \T_0,\R^*_R \R_0\right)$.
The equation is further manipulated by expressing the temperature derivative 
on the left-hand side though the mass-conservation equation, where the
density is expressed as  $\R^*=\R^*(\T^*,p^*)$, eq.~\eqref{eq:dstate},
\begin{align}
\notag
\frac{1}{\R_0}\left(\md{\R}\right)_0&=
\left(\frac{\theta^*_R}{\R^*}\pd{\R^*}{\T^*}_{|_{p^*}}\right)_{\substack{p^* = p^*_R p_0\\ \theta^* = \theta^*_R \theta_0}}
\left(\md{\T} \right)_0 +
\left(\frac{p^*_R}{\R^*}\pd{\R^*}{ p^*}_{|_{\T^*}}\right)_{\substack{\T^* = \T^*_R \T_0 \\ p^* = p^*_R p_0}} \left(\md{ p}\right)_0\\
&= -  \alpha_0 \left(\md{\T}\right)_0+ \beta_0 \left(\md{ p}\right)_0  = - \div \vec{u}_0 \, ,
\label{eq:mass_II}
\end{align}
where the dimensionless thermal expansion and isothermal compressibility 
coefficients, $\alpha_0(\theta_0,p_0)$ and $\beta_0(\theta_0,p_0)$, respectively, 
are implicitly defined by comparing the second and the third member of the equation. 
Substituting the material derivative of temperature from equation~\eqr{mass_II} 
into \eqr{ener_evol_VI} yields
\begin{align}
\R_0 \frac{c_{v0}^*}{{\cal R}_m^*} \frac{\beta_0}{\alpha_0} \md{p_0}+
\frac{c_{v0}^*}{{\cal R}^*_m}\frac{\R_0}{\alpha_0}\div\vec{u}=
-Z \T_0 \left.\pd{p_0}{\T_0}\right|_{v_0} \,\div \vec{u_0}
 + \frac{c_{p0}^*}{{\cal R}_m^*}\frac{1}{\Rey\,\Pran}\div\left(k_0 \grad\T_0\right)\, ,
\label{eq:ener_evol_VII}
\end{align}
where the zero-th order contribution to the pressure term depends only on time, 
$D p_0/Dt=d p_0/dt$, as shown in eq.~\eqref{grad_pres}. The velocity divergence then reads
\begin{align}
\div\vec{u_0}= \frac{\DS- \R_0 \frac{c_{v0}^*}{{\cal R}_m^*} \frac{\beta_0}{\alpha_0} \td{p_0}
{t}+ \frac{c_{p0}^*}{{\cal R}_m^*}\frac{1}{\Rey\,\Pran}\div\left(k_0 \grad\T_0\right)}{\DS 
\frac{c_{v0}^*}{{\cal R}_m^*}\frac{\R_0}{\alpha_0}+Z \T_0 \left.\pd{p_0}{\T_0}\right|_{\rho_0}} \ .
\label{eq:divu}
\end{align}
Summarizing, the complete system in the zero-th order approximation is
\begin{align}
\label{eq:cont}
&{\DS \pd{\R_0}{t}} + \div (\R \vec{u})_0 = 0  \\
\notag
\\
\label{eq:mom}
&\DS{\pd{(\R \vec u)_0}{t}} + \div [(\R \vec u)_0\otimes \vec u_0] =
\frac{1}{\Rey}\div \vec \Sigma_0 - \grad p_2 +\frac{1}{Fr^2} \R_0 \vec e_z  \\
\notag
\\
\label{eq:ener}
& \div\vec{u_0}=\frac{\DS -\R_0 \frac{c_{v0}^*}{{\cal R}_m^*} \frac{\beta_0}{\alpha_0} \td{p_0}
{t}+ \frac{c_{p0}^*}{{\cal R}_m^*}\frac{1}{\Rey\,\Pran}\div\left(k_0 \grad\T_0\right)}{\DS 
\frac{c_{v0}^*}{{\cal R}_m^*}\frac{\R_0}{\alpha_0}+Z \T_0 \left.\pd{p_0}{\T_0}\right|_{\rho_0}} \\
\notag
\\
\label{eq:species}
& \DS{\pd{(\R_0 Y_0)}{t}} + \div (\R_0 {\vec u}_0 Y_0) = \frac{1}{\Rey Sc} \div ({\cal D} \grad Y_0)  \\
\notag
\\
\label{eq:state}
&p_0(t) =p\left[\T_0({\bf x},t),\R_0({\bf x},t)\right]\,.
\end{align}
The crucial features of the system of equations we arrived at are worth begin emphasized: 
i) Zero-th order density $\R_0({\bf x},t)$ and temperature $\T_0({\bf x},t)$ are not 
independent fields since they are locally coupled through the equation of state 
\eqref{eq:state}. Indeed the time evolution of the  thermodynamic pressure $p_0(t)$ 
follows by integrating \eqref{eq:ener} over the, generally time dependent, flow domain 
${\cal D}(t)$ and accounting for the boundary conditions on the normal velocity component 
and on the temperature,
\begin{align}
\label{equazione_pressione}
\td{p_0}{t} \int_{{\cal D}(t)}
\R_0 \frac{c_{v0}^*}{{\cal R}_m^*} \frac{\beta_0}{\alpha_0} dV
= &
\int_{\partial {\cal D}(t)}
\frac{c_{p0}^*}{{\cal R}_m^*}\frac{1}{\Rey\,\Pran}k_0 \frac{\partial \T_0}{\partial n} dS
- \int_{{\cal D}(t) } \frac{1}{\Rey\,\Pran}k_0 \grad\T_0 \cdot \nabla \left(  \frac{c_{p0}^*}{{\cal R}_m^*} \right)dV
\\
&
- \int_{{\cal D}(t) }\left(\frac{c_{v0}^*}{{\cal R}_m^*}\frac{\R_0}{\alpha_0}+Z \T_0 \left.\pd{p_0}{\T_0}\right|_{\rho_0} \right) dV 
\int_{\partial {\cal D}(t) }\vec{u_0}\cdot {\vec n} dS \ .
\notag
\end{align}
Stepwise integration allows to determine $p_0(t)$ and to eliminate either density 
or temperature in favour of the other field through \eqref{eq:state}.
In particular,  the pressure is constant for the present problem in an unbound domain, 
i.e. $p_0(t) = p_0 = const$. ii) Acoustic waves are implicitly filtered out of the system, since 
the equation of state for pressure only enters at the lowest order, where pressure is 
spatially constant. The thermodynamic pressure is decoupled in this way from turbulence- 
and thermal-induced spatial variations of density, preventing acoustic waves from propagating 
in the medium. These effects can possibly be consistently recovered at next orders 
in the approximation. iii) The fluid density is allowed to change both in space and time 
to comply with thermal expansion effects-- heat release due to combustion and heat 
transfer are perfectly consistent within the assumed approximation limits. 
iv) No assumption is made as concerning the thermodynamic model of the fluid. This 
is essential in view of our present aim  of modelling turbulent mixing of slightly 
supercritical fluids. For the sake of definiteness, it should be stressed that resonance 
effects associated with thermo-acoustic coupling cannot be consistently dealt with within 
the range of validity of the present approximation, since, by non-linear coupling, they 
bring acoustics in foreground calling for a complete modelling of wave propagation.

\section{Thermodynamic assumptions}
\label{sec:VanderWaals}

As anticipated in the \S~\ref{sec:crit_jet}, the perfect gas equation of state is not suitable to 
describe the thermodynamic behavior close to the critical point.  
Indeed the Van der Waals equation of state is presumably the simplest extension of the 
perfect gas model able  to consistently deal with the thermodynamics of a diatomic gas at 
a  slightly super-critical pressure. It  will be  hereafter assumed for its simplicity in the 
present application of  the Low-Mach number expansion which, however, can be easily 
adapted to more general cases, like e.g. the Peng-Robinson  equation of 
state~\cite{penrob}, see Appendix~\ref{sec:peng_robinson}.

The Van der Walls  theory is directly derived from  statistical mechanics, see e.g. Ref~\onlinecite{zhu}, and 
allows a complete and straightforward thermodynamic characterization of the gas starting 
from the Helmholtz free energy $f^*  = - N k_b \theta \ln\left( \cal Z^*\right)$ of 
the system expressed in terms of the canonical partition function ${\cal Z}^*(N,V,\theta)$ 
of the corresponding atomistic model.  Here $N$ is the number of gas molecules, $V$ 
the volume and $k_B$ the Boltzmann constant, see Appendix~\ref{sec:appendix VdW} for a 
brief review of the subject. The pressure equation of state follows as
\begin{align}
p^*=-\left.\pd{f^*}{V^*}\right|_{\T^*,N}
=\frac{{\cal R}_m^* \T^* \R^*}{1-b'^*\R^*}-a'^*\R^{*2}\, ,
\label{eq:p_rho}
\end{align}
where $a'^*$ and $b'^*$ are the Van der Waals coefficients that
account for intermolecular forces and  excluded volume, respectively.
Thermal expansion and  isothermal compressibility coefficients, $\alpha^*$ 
and $\beta^*$ respectively,
\begin{align*}
\alpha^*=-\frac{{\cal R}_m^*}{2a'^*\R^*-
3a'^* b'^*\R^{*2}- p^*b'^*-{\cal R}_m^*\T^*}\\
\beta^*=\frac{\left(1-b'^*\R^*\right)/\R^*}
{2a'^*\R^*-3a'^*b'^*\R^{*2}- p^*b'^*-{\cal R}_m^*\T^*}
\end{align*}
follow directly from the pressure equation of state. All these thermodynamic 
relationships can be expressed in dimensionless form.  
Assuming $p = p_0 + {\cal O}\left(\widehat{Ma}^2 \right)$, with similar expressions for $\theta$ and $\rho$, leads to the
leading order contributions in the  Low Mach number expansion for pressure, thermal expansion and isothermal compressibility,
\begin{align}
\label{eq:p_rho_adim}
p_0&=\frac{ \T_0 \R_0}{Z (1-b'\R_0)}-a'\R_0^2\, ,\\
\alpha_0&=-\frac{1}{Z\left(2a'\R_0-3a'b'\R_0^2- p_0b'\right)-\T_0}\\
\beta_0&=\frac{\left(1-b'\R_0\right)/\R_0}{2a'\R_0-3a'b'\R_0^2- p_0b'-\T_0/Z}
\end{align}
where  $a'=a'^* \R_R^{*2}/p_R^*$ and  $b' = b'^*\R_R^*$. 

Applying the analysis presented in  section \S~\ref{sec:method}
to the Van der Waals equation of state, most of the equations in system 
\eqr{cont}-\eqr{state} remain unchanged, while velocity divergence and pressure 
equation of state become
\begin{align}
\label{eq:ener_last}
\div {\vec u}_0 & ={\displaystyle \frac{1}{Z \,p_0}} \left[ \frac{c_{p0}^*}{{\cal R}_m^*}
\frac{1}{\Rey \Pran}\div(k\grad \T)_0\right]
\left[\frac{1}{\displaystyle{1+\frac{\left(\gamma-2\right) a \R_0^2}
{p_0 \gamma}+\frac{2 a'b' \R_0^3}{p_0 \gamma}}} \right]\\
\notag
\\\label{eq:state_last}
\T_0 & =Z\,{\displaystyle \frac{p_0}{\R_0}\left(1+a' \frac{\R_0^2}{p_0}\right) 
\left(1-b' \R_0\right)}\, .
\end{align}
In eq.~\eqref{eq:ener_last}  $\gamma = c_{p0}^*/c_{v0}^*$ and the spatially 
constant pressure has been assumed constant also in time, $p_0(t) = {\rm const}$,
as appropriate for  a  free jet configuration where the discharge pressure is a known parameter.
Overall, the system is formed by seven equations in the seven unknowns  $\R_0$, 
${\vec u}_0$, $p_2$, $Y_0$ and $\T_0$.
Peculiar feature is the presence of the hydrodynamic pressure $p_2$ 
in the momentum equation \cite{majset}. Under this respect, 
the system is similar to the incompressible Navier-Stokes 
equations where the  pressure takes the mathematical role of a Lagrange 
multiplier that allows to enforce the constraint~\eqr{ener_last} 
on the velocity.

As a final comment on the mathematical model, we like stress once more that the same 
machinery would work in the same way  also for different gas models. 

\section{Numerics and physical parameters}
\label{sec:variable_density_jet}

\begin{figure}
\centering
\includegraphics[width=.5\textwidth]{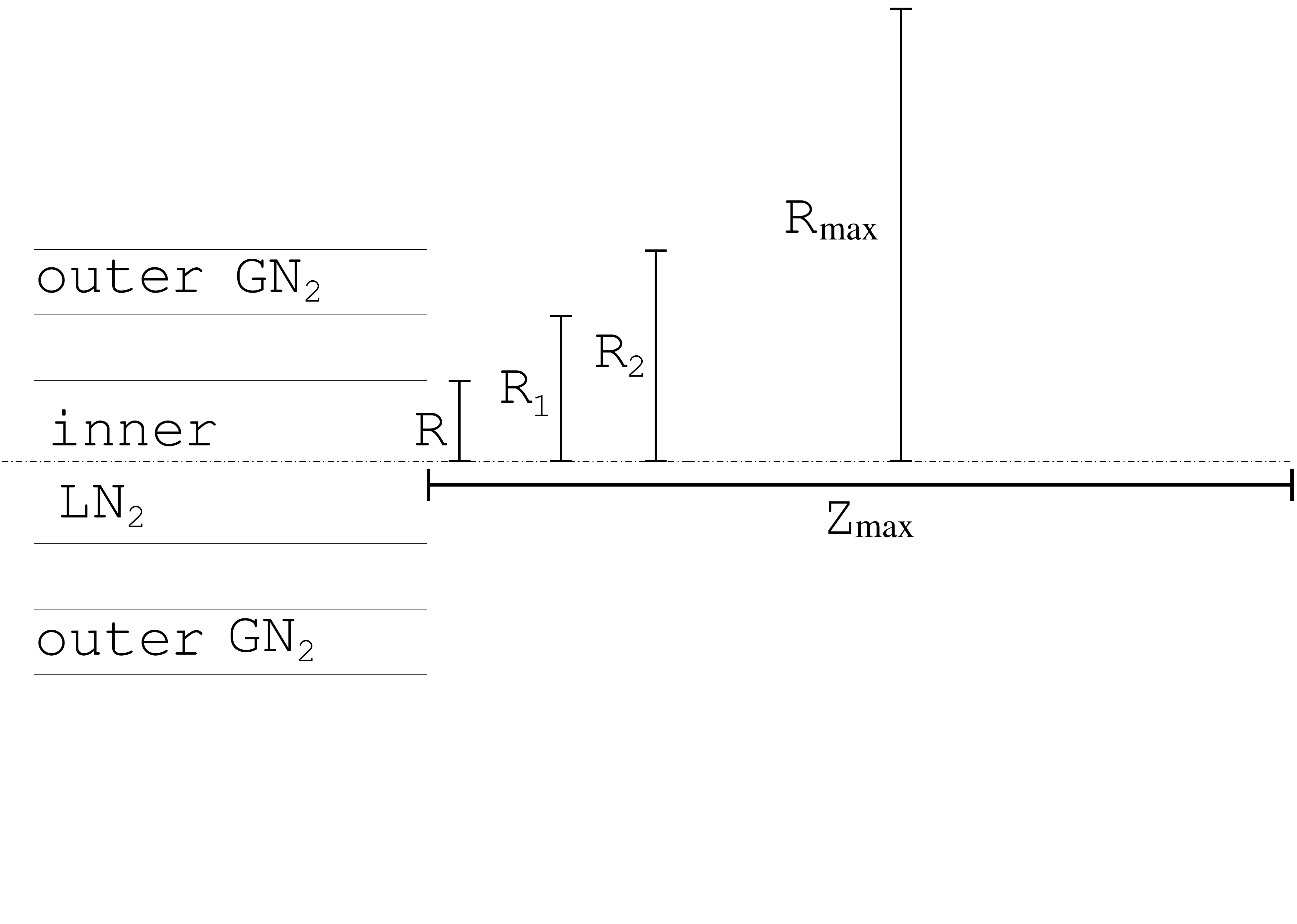}
\caption{\label{geom_coax} Schematic diagram of the coaxial 
jet used in the present simulations. Liquid-like density Nitrogen $N_2$ is injected through 
the ``Inner'' nozzle while the ``Outer'' nozzle discharges gaseous Nitrogen $N_2$ in the high pressure
environment. In dimensionless variables the inner radius is $R=1$.}
\end{figure}

The DNS of a coaxial jet of a Van der Waals gas --hereafter  referred to as real-gas--
is performed employing  cylindrical coordinates
$\left(r, \varphi, z \right)$, with $z$ the axial coordinate, $r$ the radial coordinate in the transverse plane and
$\varphi$ the angle, in  the cylindrical  domain with dimensions $[\varphi_{max}\times R_{max}\times Z_{max}]
=[2\pi\times 18 R\times 20R]$ with $[N_\varphi\times N_R\times N_Z]
=[128 \times 281 \times 600]$ collocation points.
The geometry of the system -- figure~\ref{geom_coax} -- consists of a coaxial jet 
with inner nozzle radius $R$ and with  inner and outer radii of the  outer  jet
$R_1 = 1.2R$ and $R_2 = 1.5R$, respectively. The jet discharges in a cylindrical 
domain with radius $R_{max} = 18 R$, large enough to allow the use of
traction free  conditions on the side boundary. The axial extent of the computational domain is $Z_{max}= 20 R$ 
with convective~\cite{orl} boundary condition used at the exit 
section. The numerics models an apparatus with sufficiently long inlet manifold to 
have fully developed turbulence at the core inlet. On the contrary, the outer stream 
is considered to be fed by a short annular manifold, such that the inflow velocity is 
constant through the section. The core inlet turbulence is taken from a companion 
turbulent pipe flow at matching conditions, see e.g. Refs~\onlinecite{piccas,picsarguacas} for more details. 
In the different simulations addressed 
below the mean inlet velocity and density  typically change, and the different cases 
are compared at the same turbulent intensity $q/|\vec{u}^*_R|$, with
$q = \sqrt{\langle u'_i u'_i \rangle}$ and $u'_i$ the instantaneous velocity 
fluctuation.

In the radial direction  grid stretching is applied to resolve the shear layer occurring 
at the boundary between the internal and the external jet and between the external 
jet and the surrounding environment. Given the Kolmogorov scale at the inlet of the 
core jet ($\eta\simeq 0.01\,R$) the grid size, $\Delta r=0.0125\,R$  in the shear 
layers, is able to accurately capture the finest scales of turbulence. The present 
discretization also enables capturing the strong density gradients which take place 
across the inner shear layer separating the outer stream from the core jet, see also Refs
\onlinecite{picbattrocas,batpictrocas} for similar considerations in the context of
combustion.
The basic simulation to be discussed deals with a real gas coaxial Nitrogen jet in  
transcritical conditions, to be addressed as simulation A to distinguish it  from several 
others we performed to highlight real gas effects on turbulent jet dynamics. Geometry and  
thermodynamic parameters are chosen to be similar to the coaxial jet experiments presented 
by \emph{Mayer et al.}. Clearly the Reynolds number amenable to DNS is 
significantly smaller than the experimental one, $Re_D = 6000$ compared to the 
experimental value of order of ${\cal O}(10^4-10^5)$~\cite{mayschschsch,schrodleycan}.
This  Reynolds number value   can be achieved, e.g., by considering an inner nozzle of  
diameter $D^* =1.8 \, {\rm mm}$ with typical injection velocity 
$|\vec{u}^*_{core}| = 0.2  \, {\rm m/s}$, density $\R^*_{core}  = 320 \, {\rm kg/m^3}$ and 
dynamic viscosity $\mu^* = 1.9 \times 10^{-5}  \, {\rm Pa/s}$, corresponding to slightly 
supercritical Nitrogen. The sound speed at core inlet is $c^*  = 179 \, {\rm m/s}$, leading to 
a Mach number $Ma = 1 \times 10^{-3}$ consistently with the adoption of the low Mach 
number expansion described in the previous sections.

The Nitrogen coaxial jet  is assumed to discharge in a gaseous $N_2$ environment at 
$p^*_{env} = 4.0\, MPa$, slightly above the Nitrogen critical pressure 
($p_c^*\simeq3.4MPa$,  $p^*_{env}/p_c^* = 1.168$).  The core jet is injected at a density 
slightly exceeding the critical density $\R_{core}^*\simeq1.04 \R_c^*$, while the  external 
stream has density ten times lower, $\R^*_{core}/\R^*_{ext}=10$, matching  that of the 
surrounding environment $\rho_{core}^*>\rho_ {c}^*>\rho_{ext}^* = \rho^*_{env}$.
The injection temperature of the core  is $\T^*_{core} = 131.46 \,{\rm K}$ (with $\T^*_c=126.2\, 
{\rm K}$), while external jet and environment are at $\T^*_{env} = 529.79 \, {\rm K}$, with 
a temperature ratio $\T^*_{env}/\T^*_{core} \simeq 4$.
\begin{figure}
\centering
\includegraphics[width=.95\textwidth]{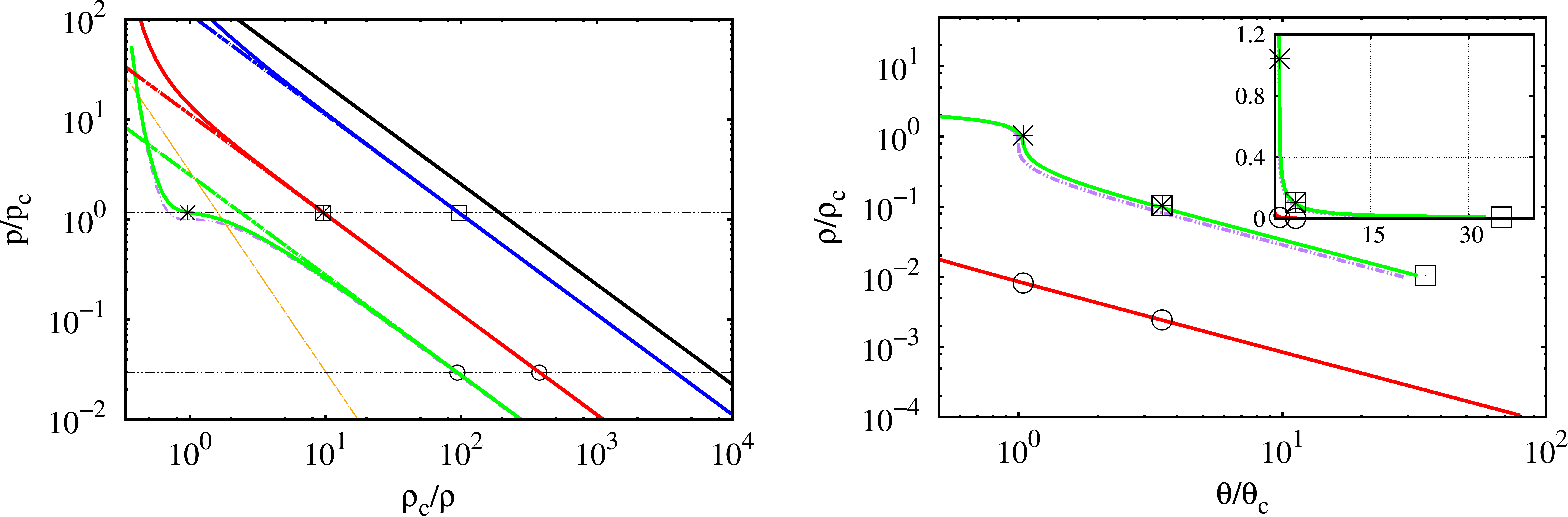}
{\scriptsize \put(-445,150){\bf (a)}}
{\scriptsize \put(-210,150){\bf (b)}}
\caption{\label{fig:P-V_diagram}
Panel (a): pressure-density diagram (log-log coordinates) for the Van der Waals equation of state. 
Solid lines: isotherms at different temperatures for the Van der Waals equation of state.
Dash-dotted lines: isotherms for the perfect fas equation of state. 
Dotted line (purple in the electronic version): critical isotherm.
Symbols: thermodynamics injection conditions of each simulation, namely Sim A and C (asterisks), 
Sim B (squares), and Sim D (circles).
Panel (b): ${\hat \rho}-{\hat \theta}$ phase diagram and  injection conditions 
of each simulation, same symbols as in panel (a). Inset: same quantities in linear scale.
}
\end{figure}

Panel (a) of figure~\ref{fig:P-V_diagram} provides the pressure-density diagram 
(log-log coordinates) for the Van der Waals equation of state in reduced variables 
${\hat \theta} = 3/8 \left({\hat p} + 3 {\hat \rho}^2 \right) \left(1/{\hat \rho} -1/3 \right)$ 
and $\hat p = p/p_c$, $1/\hat \rho = \rho_c/\rho$, ${\hat \theta} = \theta/\theta_c$. 
The dotted line is the critical isotherm ${\hat \theta} = 1$. The other solid lines are 
isotherms at increasing temperature. The dash-dotted lines sketch the corresponding isotherms 
for the perfect gas equation of state,  ${\hat \theta} = 3 {\hat p}/\left(8 {\hat \rho}\right)$.
The limit behavior of the perfect gas is achieved when two conditions are met, 
namely i) ${\hat p} \gg 3 {\hat \rho}^2$, with the curve ${\hat p} = 3 {\hat \rho}^2$ shown as the 
dash-double-dotted line of slope $-2$, and ii)
${\hat \rho} \ll 3$, which is the vertical axis 
delimiting the plot on the left.
Increasing the temperature the perfect gas limit is recovered, uppermost solid line 
of slope $-1$.  Moving along an isotherm the perfect gas limit is reached as well 
at sufficiently low pressure.

The symbols reported in panel (a) of figure~\ref{fig:P-V_diagram} provide the thermodynamic conditions for 
the simulations considered in the paper. The two asterisks are the working points 
for core jet and environment (leftmost and rightmost symbol, respectively) for the 
real gas cases, simulations $A$ and $C$ of Table~\ref{tab:sim_cond}. 
The two open squares (one of which superimposed to an asterisk) give the corresponding 
points for the perfect gas simulation, simulation $B$ in the same Table. Both 
asterisks and open squares are on the same isobar, at slightly supercritical 
pressure ${\hat p} = 1.17$. The density ratio for simulations A, C (asterisks) 
and B (open squares) is the same, $\rho_{core}/\rho_{ext}  \simeq 10$.
On the two isotherms concerning the real gas conditions (asterisks) two additional 
working points at a substantially lower pressure, ${\hat p} = 0.02941$
corresponding to atmospheric pressure for Nitrogen, are denoted by  open circles 
and provide the injection conditions for simulation D. Here the gas  behavior can 
be approximated with the perfect gas law. The temperature ratio for simulation D 
is the same as that for simulations A and C. Clearly the density ratio is instead 
much lower, $\rho_{core}/\rho_{ext}  \simeq 4$. 
The ${\hat \rho}-{\hat \theta}$ phase diagram of the conditions 
of each simulation is represented in panel (b) of figure~\ref{fig:P-V_diagram}.

The two constants of the Van der Waals model, $a^* = 3 p^*_c /\R_c^{*2}$ and $b^* =1/\left(3 
\R^*_c \right)$, are determined from the critical pressure and density of Nitrogen.
Assuming the value of the universal gas constant, the critical temperature is estimated as 
$\T^*_c = 8 p^*_c/\left(3 {\cal R}^*_m   \R^*_c\right) = 97.22 \, {\rm K}$ 
in comparison with  $126.2$  pertaining to  actual Nitrogen.
For our purposes, the Van der Waals model reproduces acceptably well the behavior of 
Nitrogen with the advantage of having a clear and reasonably simple theoretical 
derivation. In case better accuracy were needed, alternatives are available, 
e.g.\ the Peng-Robinson model. 

For the reader's convenience, it may be worth mentioning what the rationale behind the parameters selection for  simulation B is.
Simulation B,  squares in figure~\ref{fig:P-V_diagram}, has same pressure and density ratio 
$\R^*_{core}/\R^*_{env} = 10$ of simulation A. The difference is the larger injection 
temperatures such that Nitrogen behaves as a perfect gas, $\T^*_{core} = 529.79 \, {\rm K}$ 
and temperature ratio $\T^*_{env}/\T^*_{core} =  \R^*_{core}/\R^*_{env} = 10$.
At constant pressure, the dynamics of a strongly subsonic 
perfect gas is substantially controlled by density ratio, Prandtl and Reynolds number. 
Instead, the parameters entering the description of  real gas dynamics  also include the 
distance of the injection conditions from the critical point.
Indeed simulation B is designed to achieve the same density ratio and injection pressure 
of simulation A, using the perfect gas equation of state in the region of the parameter 
space where Nitrogen recovers the perfect gas behavior.

Indeed, a part from the real vs perfect gas issue, an additional difference exists between  
simulations A and B, namely the  different range of  temperatures which affects the 
transport coefficients. In order to discriminate between the effect of  thermodynamic 
behavior and temperature range, a third simulation, C,  has been conceived with same 
thermodynamic  conditions and gas model of simulation A (real gas), now artificially 
enforcing the Prandtl number of simulation B (perfect gas).

It is important to recall that in all the three cases just considered the density ratio between 
inner core and external jet plus environment is a large one, $\R^*_{core}/\R^*_{env} = 10$. 
It is worthwhile comparing the results with a fourth simulation, D, where the density ratio 
is substantially lower due to a lower environment pressure, as for Nitrogen at atmospheric 
pressure and same injection temperatures of the basic simulation $A$ (real gas),
$\T^*_{core} =131.46  \, {\rm K}$,  $\T^*_{env}/\T^*_{core} \simeq 4$, which results in 
$\R^*_{core}/\R^*_{env} = 4$. This simulation is performed with the Van der Waals 
equation of state and the actual transport coefficients of Nitrogen in the parameter region 
where Nitrogen behaves almost like a perfect gas (i.e.\ the real system  could have been 
accurately approximated with the perfect gas equation of state).
\begin{table}
\centering
\begin{tabular}{c|ccc}
sim A-C        & inner jet & outer jet & surrounding environment\\ \hline
$p^*$ &$p^*_R$&$p^*_R$&$p^*_R$\\
$\R^*$  & $\R^*_R$  & $0.1\, \R^*_R$&$0.1\, \R^*_R$\\
$\T^*$& $\T^*_R$ &$4\,\T^*_R$& $4\,\T^*_R$\\
$(\R^* U^* A^*)_{inj}$ & $\R^*_R |\vec u_R^*| \pi R^{*2}$ & $0.234 \,\R^*_R |\vec u_R^*| \pi R^{*2}$&-- \\
$(\R^* U^{*2} A^*)_{inj}$ & $\R^*_R |\vec u_R^*|^2 \pi R^{*2}$ & $1.296 \,\R^*_R |\vec u_R^*|^2 \pi R^{*2}$&-- \\
\end{tabular}\\\vspace{.5cm} 
\begin{tabular}{c|ccc}
sim B        & inner jet & outer jet & surrounding environment\\ \hline
$p^*$ &$p^*_R$&$p^*_R$&$p^*_R$\\
$\R^*$  & $\R^*_R$  & $0.1\, \R^*_R$&$0.1\, \R^*_R$\\
$\T^*$& $\T^*_R$ &$10\,\T^*_R$& $10\,\T^*_R$\\
$(\R^* U^* A^*)_{inj}$ & $\R^*_R |\vec u_R^*| \pi R^{*2}$ & $0.234 \,\R^*_R |\vec u_R^*| \pi R^{*2}$&-- \\
$(\R^* U^{*2} A^*)_{inj}$ & $\R^*_R |\vec u_R^*|^2 \pi R^{*2}$ & $1.296 \,\R^*_R |\vec u_R^*|^2 \pi R^{*2}$&-- \\
\end{tabular}\\\vspace{.5cm} 
\begin{tabular}{c|ccc}
sim D        & inner jet & outer jet & surrounding environment\\ \hline
$p^*$ &$p^*_R$&$p^*_R$&$p^*_R$\\
$\R^*$  & $\R^*_R$  & $0.25\, \R^*_R$&$0.25\, \R^*_R$\\
$\T^*$& $\T^*_R$ &$4\,\T^*_R$& $4\,\T^*_R$\\
$(\R^* U^* A^*)_{inj}$ & $\R^*_R |\vec u_R^*| \pi R^{*2}$ & $0.234 \,\R^*_R |\vec u_R^*| \pi R^{*2}$&-- \\
$(\R^* U^{*2} A^*)_{inj}$ & $\R^*_R |\vec u_R^*|^2 \pi R^{*2}$ & $1.296 \,\R^*_R |\vec u_R^*|^2 \pi R^{*2}$&-- \\
\end{tabular}
\caption{\label{tab:sim_cond} Thermodynamic and physical conditions at
injection and surrounding environment normalized with the reference quantities. 
Top table: real gas  (simulations A-C) where the reference quantities
are: $p^*_R=1.178 p^*_c$, $\R^*_R=1.0424\R^*_c$ and $\T^*_R=1.0417\T^*_c$.
Middle table: perfect gas  (simulation B) with reference quantities:
 $p^*_R=1.178 p^*_c$, $\R^*_R=0.10424 \R^*_c$ and $\T^*_R=4.198 \T^*_c$.
Bottom table: real gas  (simulation D)  where the reference quantities
are $p^*_R=0.0294117p^*_c$, $\R^*_R=0.010672 \R^*_c$ and $\T^*_R=1.04\T^*_c$; due to 
low  pressure and density the limit of perfect gas behavior is almost reached.
}
\end{table}

Concerning momentum, in all the four cases the momentum ratio is kept constant at
$\R^*_{ext} |\vec u^*_{ext}|/\left(\rho^*_{core} |\vec u^*_{core}| \right) = 0.4$.
For the purpose of making equations dimensionless reference conditions are selected as the 
corresponding  core jet features 
$\R^*_R = \R^*_{core}, p^*_R = p^*_{core }=p^*_{env}, |\vec u^*_R| = |\vec u^*_{core}|$.
\begin{figure}
\centering
\includegraphics[width=.95\textwidth]{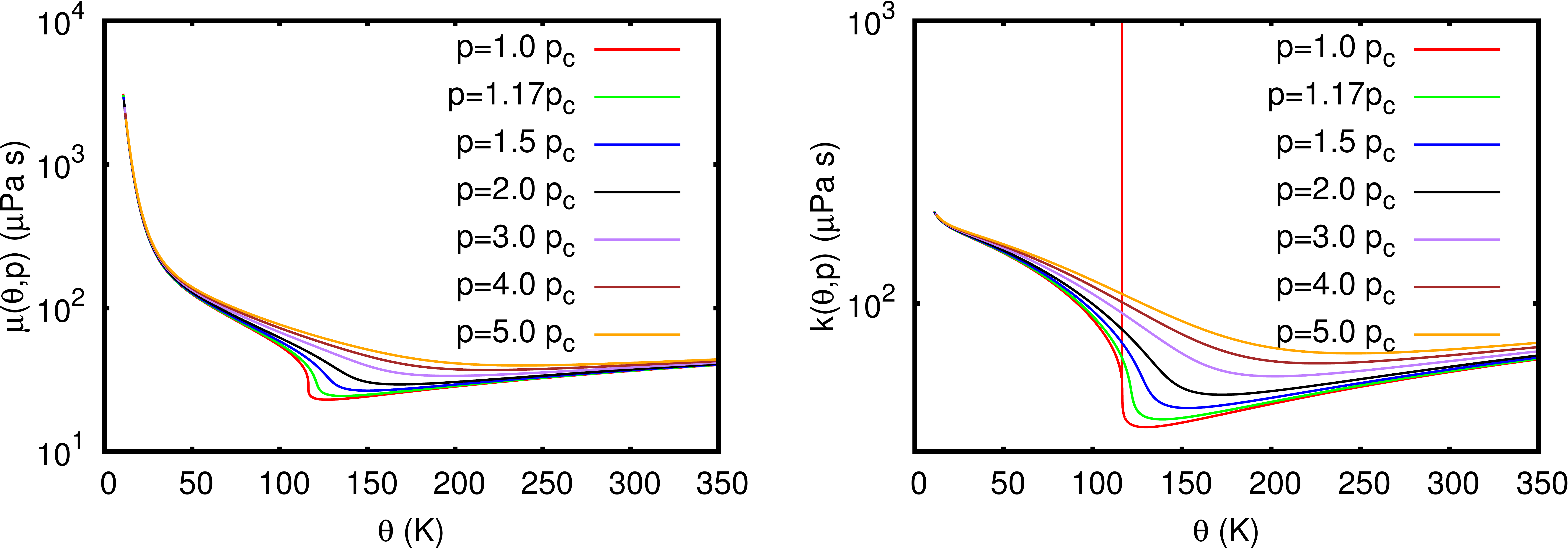}
{\scriptsize \put(-455,150){\bf (a)}}
{\scriptsize \put(-220,150){\bf (b)}}
\caption{\label{fig:visc_ther} Dependence of Nitrogen viscosity $\mu(\T,p)$, panel (a),
and thermal conductivity $k(\T,p)$, panel (b), on temperature and pressure. 
The complete analytical relations are provided in Appendix~\ref{sec:visc_ther}.}
\end{figure}
Table~\ref{tab:sim_cond}  provides detailed  information for the different 
simulations. We stress once more that in cases $A$ and $B$  density and momentum 
ratios between core and outer stream are the same,  $\rho^*_{core}/\rho^*_{ext} = 10$ and 
$\left(\R^*_{ext}|\vec u^*_{ext}|\right)/\left(\R^*_{core}|\vec u^*_{core}|\right)= 0.4$.
The comparison is aimed at addressing the effects of supercritical injection on jet 
dynamics and mixing process, at low-Mach number, hence with neglecting acoustic effects. 
The two simulations mainly differ for the temperatures of core and outer jet.  While in  
the perfect gas case, for given pressure, the density ratio $\rho^*_{core}/\rho^*_{ext}=10$ 
results in the temperature ratio $\T^*_{core}/\T^*_{ext}=0.1$, in the real gas case the 
temperature ratio is  $\T^*_{core}/\T^*_{ext}\simeq0.25$. The bottom panel of 
Table~\ref{tab:sim_cond} provides the physical and thermodynamic conditions of 
simulation D. Here since the temperature ratio $\T^*_{ext}/\T^*_{core}$ matches that of 
simulation A, the density ratio is much smaller than all the other cases,
$\R^*_{core}/\R^*_{ext}=\T^*_{ext}/\T^*_{core}\simeq 4$.

The model is completed with the expressions for dynamic viscosity and thermal 
diffusivity as a function of pressure and temperature. We use here the relations provided 
in Ref~\onlinecite{lemjac},
\begin{equation}
\label{eq:viscosity&thermal_diffusivity}
\begin{array}{cl}
\mu^*&=\mu^{*0}(\T^*)+\mu^{*r}(\tau,\delta)\\
&\\
k^*&=k^{*0}(\T^*)+k^{*r}(\tau,\delta)+k^{*c}(\tau,\delta)
\end{array}
\end{equation}
where $\tau= \T^*_c/\T^*$ and $\delta^* =  \R^*/\R^*_c$, with $\mu^{*0}$ and 
$k^{*0}$ the perfect gas viscosity and thermal conductivity, respectively, 
$\mu^{*r}$ and $k^{*r}$ the so-called residual fluid contributions, and 
$k^{*c}$  the critical enhancement of thermal conductivity~\cite{lemjac}. Since the 
effect of the critical condition on viscosity is negligible no critical enhancement is 
considered. The expressions entering eqs.~\eqref{eq:viscosity&thermal_diffusivity}
are explicitly reported in  Appendix \ref{sec:visc_ther} with the model constant chosen to 
reproduce  Nitrogen. 
Figure~\ref{fig:visc_ther} shows the resulting behavior of viscosity and thermal 
conductivity as a function of temperature and pressure.

Using the core jet parameters as reference conditions,   eqs.~\eqref{eq:reference}, 
the Prandtl number, eq.~\eqref{eq:parameters}, is  $Pr=0.35$ for simulation A (real gas),  
$Pr=0.6$  for simulation B (high pressure perfect gas case) and C
(real gas with transport coefficient matching the high pressure perfect gas case B) and   
$Pr=0.75$  for simulation D (low pressure case, real gas behaving like a perfect gas).
These values are determined through the transport  coefficients 
\eqref{eq:viscosity&thermal_diffusivity}  and the specific heat coefficient \eqref{eq:cp}
evaluated at the respective reference thermodynamic conditions.  It is worth stressing 
that the difference in the Prandtl number between cases A and B/C is substantial.

In all cases buoyancy is neglected since it would have  introduced an explicit dependence 
on the density ratio, thus hampering a fair comparison between the different gas models.

\section{Results}
\label{sec:results}

\subsection{Mean fields}
\begin{figure}
\centering
\includegraphics[height=.55\textwidth]{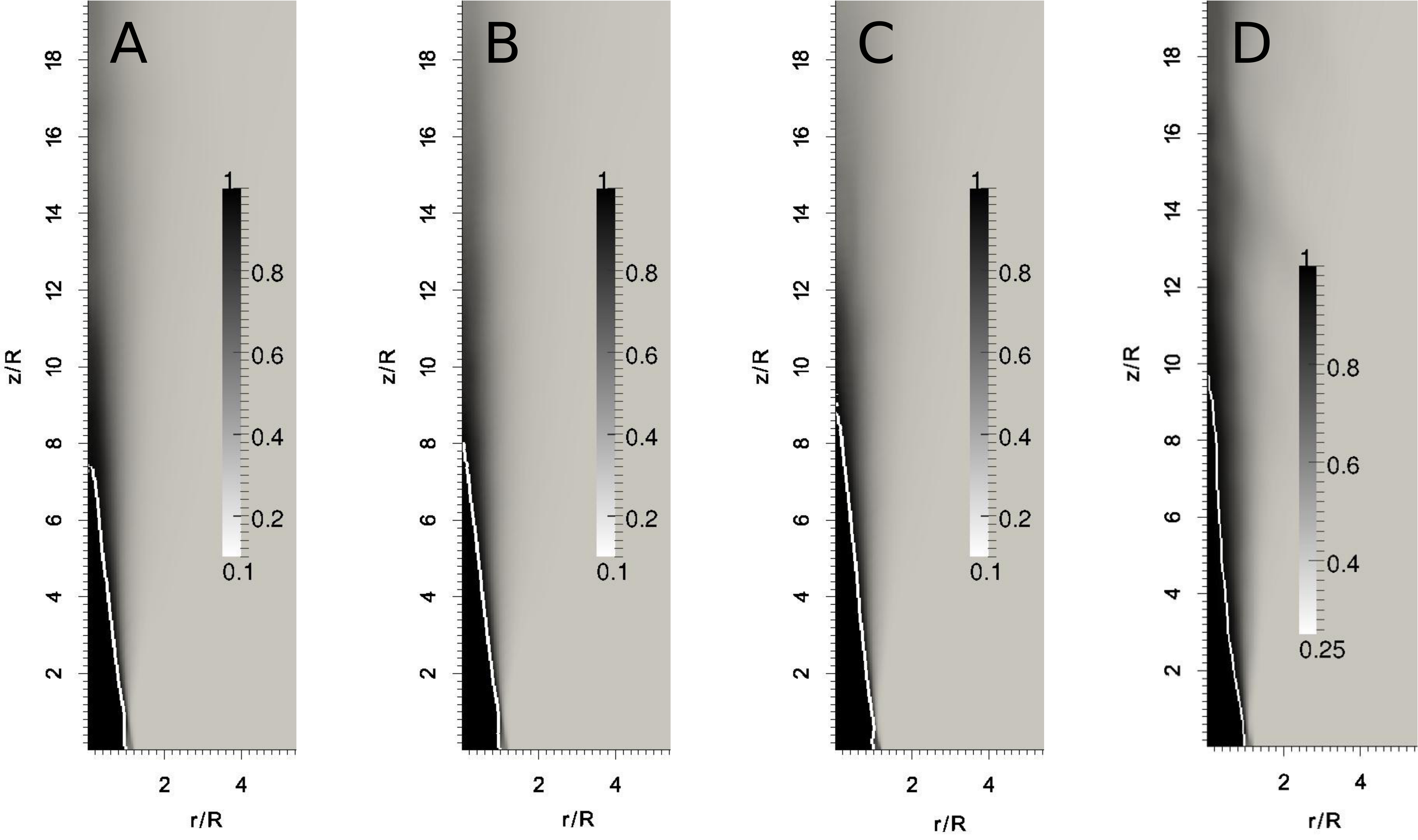}
\caption{\label{fig:mean_den} Normalized mean density fields $\R^*/\R^*_{jet}$.
From left to right: real gas jet (sim. A), perfect gas jet (sim. B), real gas jet with 
perfect gas transport properties (sim. C) and real gas jet at low pressure (sim. D), 
respectively.}
\end{figure}
As anticipated four simulations are considered. For each  of them average fields 
are extracted by ensemble averaging of about $200$ independent instantaneous configurations 
separated in time by $0.25 R/U_{r}$. Each sample was acquired after the flow reached 
statistically steady conditions. The typical correlation time $\tau_c = 0.083 R/U_r$ 
is estimated from the autocorrelation of the axial velocity fluctuation 
$u'_z(\phi,r,z,t) = u_z - \langle u_z \rangle$ on the jet axis one diameter 
downstream of the inlet section ($r = 0, z =D$), 
$$\tau_c = \frac{\displaystyle \int_0^{50 R/U_r} \int_0^{50 R/U_r} u'_z(\phi,0,D,t) u'_z(\phi,0,D,t+\tau) dt d\tau}{\displaystyle \int_0^{50 R/U_r} {u'_z}^2(\phi,0,D,t)  d \tau}\ .$$ 
Here and in the following, the subscript $0$  is dropped from the variables, 
since no confusion may arise and the hydrodynamic pressure $p_2$ is never mentioned explicitly.

The dimensionless average density $\R^*/\R^*_{core}$ is illustrated in 
figure~\ref{fig:mean_den} where  it ranges from 1 in the jet core to 0.1 in 
the outer stream  for the three density matched simulations including both the 
two real gas simulations (case A and C) and  the perfect gas simulation (case B). 
The density ranges instead from $1$ (core jet) to $0.25$ in the outer stream for 
simulation D (low pressure, real gas case) which is temperature matched with 
the high pressure, real gas simulations A and C. The  mixing efficiency may be 
quantified by the extension of the core region measured, e.g., by the intercept on the
jet axis of a selected high-density isoline,  $\R^* = 0.95 \R_{core}^*$ in the present case. 
The figure shows that the low pressure case (D) exhibits the longest core 
length.

\begin{figure}
\centering
\includegraphics[height=.6\textwidth]{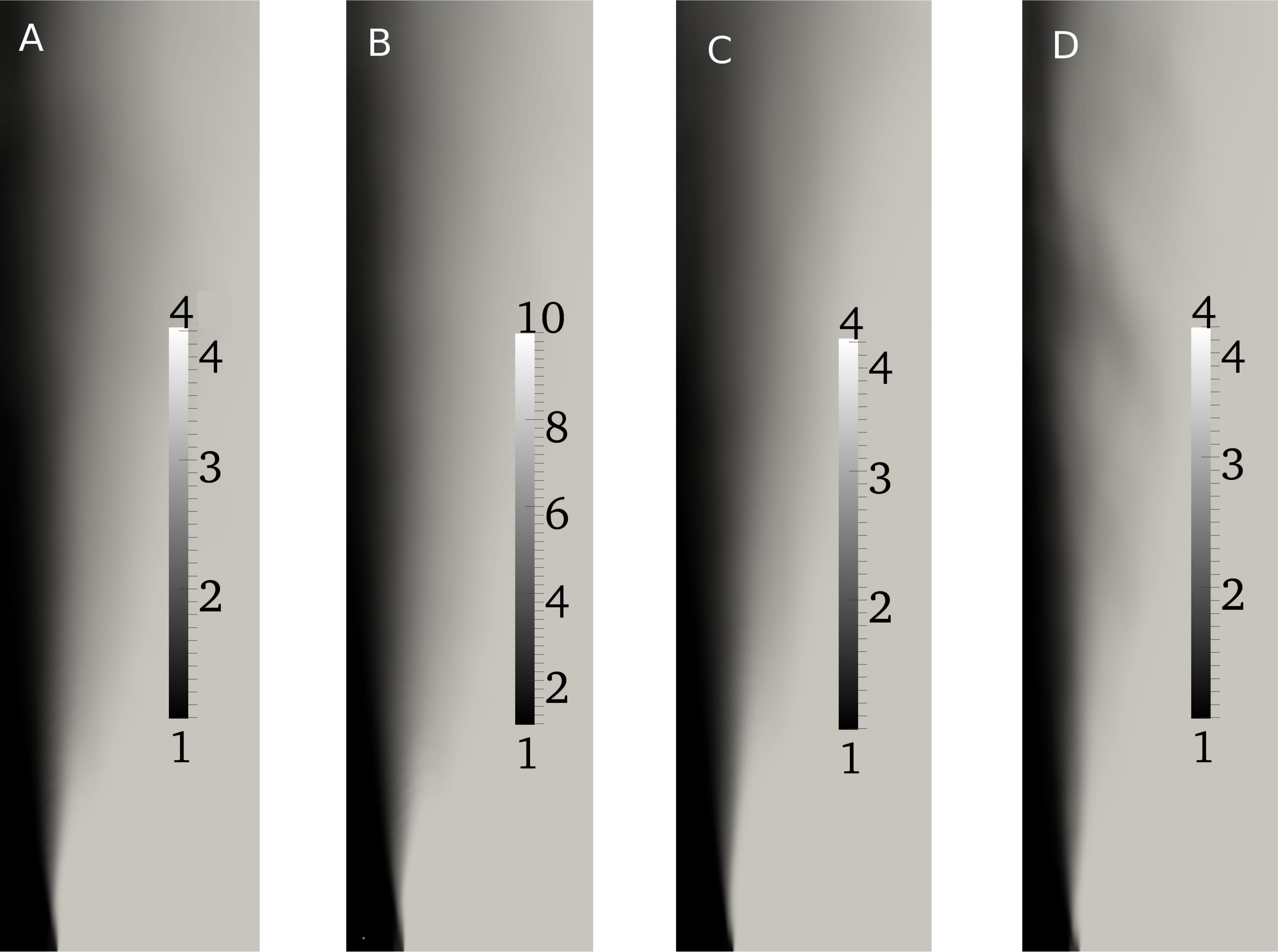}
\caption{\label{fig:mean_temp} Normalized mean temperature fields $\T^*/\T^*_{ext}$.
From left to right: real gas jet (sim. A), perfect gas jet (sim. B), real gas jet with 
perfect gas transport properties (sim. C) and real gas jet at low pressure (sim. D), 
respectively.}
\end{figure}

Concerning the dimensionless temperature  $\T^*/\T^*_{ext}$, figure~\ref{fig:mean_temp}, 
its normalized mean  field ranges from 4 in the jet core to 1 in  the outer stream for the 
real gas cases (A and C). For the perfect gas case B the temperature ranges instead 
from 10 in the jet core to  1 in the outer part. In other words, at fixed density ratio, 
a higher temperature ratio characterizes the perfect gas case as a consequence of the 
different equation of state that become crucial in near critical conditions.

\subsection{Instantaneous fields}

\begin{figure}
\centering
\includegraphics[scale=.185]{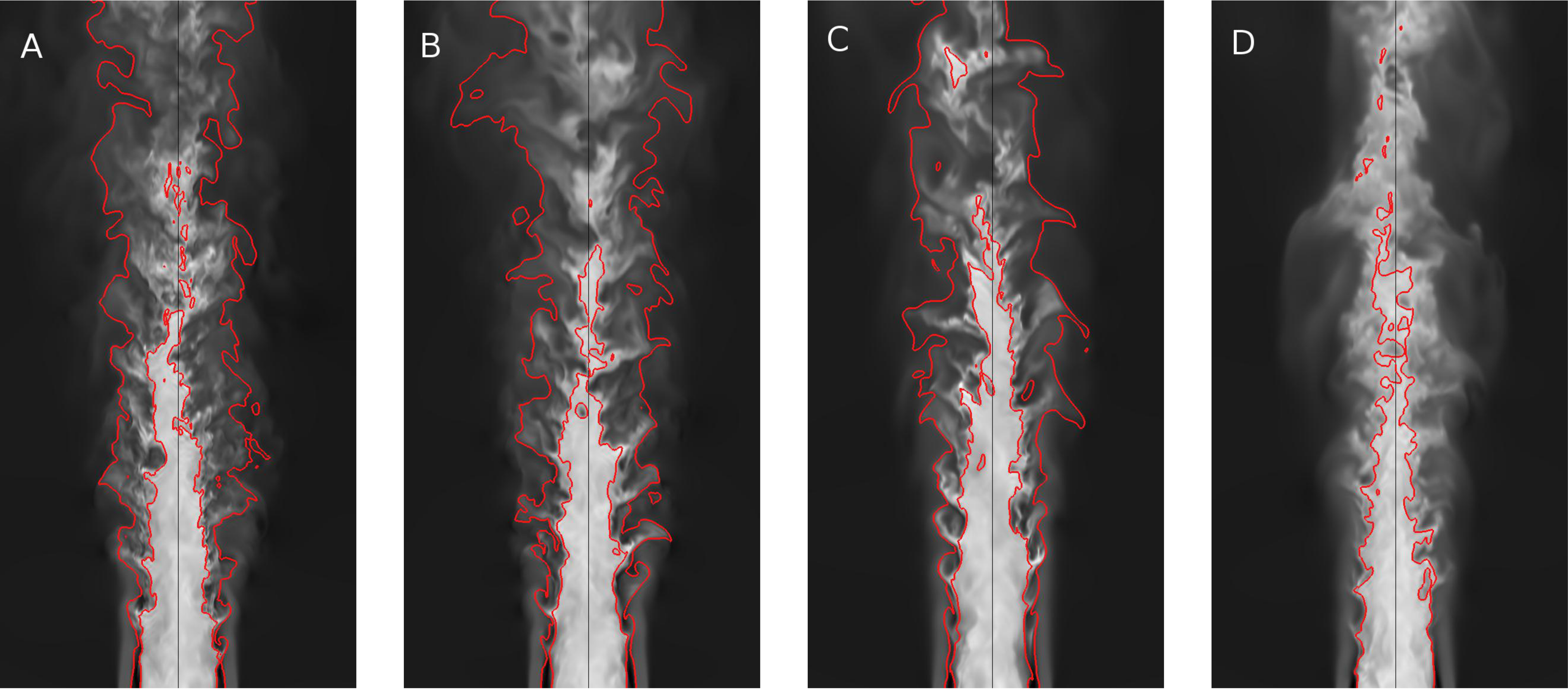}
\caption{Axial momentum field ($\rho u_z$, gray scale contours) 
with two density isolevels ($\rho =0.9, 0.2$, respectively) shown by solid 
lines.  From left to right: case A (real gas, high pressure), case B (perfect gas, 
high pressure), case C (real gas, with modified transport coefficients) and case D 
(real gas, low pressure). In the last case the $0.2$ density isoline does not exist.
\label{fig:inst}}
\end{figure}

The four panels in Figure~\ref{fig:inst} compare planar cuts of instantaneous axial 
momentum fields, $\rho u_z$, shaded contours. 
The solid lines  denote two  density isolevels, namely $\rho=0.9$ and $\rho=0.2$, respectively.
In the near field, close to the inlet, typical structures are apparent  at the shear 
layers formed between inner and outer streams where high density contrast 
occurs between inner and outer jet (cases A, B, and C). They correspond  to the 
finger-like objects  observed in the experimental 
snapshot~\cite{chetalcoy,chetalmaybrasmischosc} and known to characterize supercritical 
jets also at moderate Mach numbers. 
Such structures are much less apparent in the low density ratio case D (atmospheric 
pressure case). Interestingly they also exist in the high pressure perfect gas jet 
(B), confirming that they are associated with the high density contrast between inner 
and outer stream. The structures are similar to the ligaments occurring in the 
break-up of a liquid jet~\cite{eggvil}. They are formed through a similar kinematic 
mechanism but are  associated with different thermodynamic 
phenomenologies~\cite{chetalcoy,segpol}. The liquid (subcritical) jet is 
characterized by two immiscible phases, liquid and gaseous, with no mutual 
diffusion. In this case the joint effect of shear and capillary instability promoted by 
the surface tension acting at the liquid-gas interface induces droplet formation.
The external high velocity gas stream  stretches the liquid core, forming finger-like 
structures that elongate until they break-up into droplets due to 
capillarity~\cite{eggvil,bellan}. For the cases at high core density (A, B, 
and C), the density continuously varies from a liquid-like large value in the core to 
a gas-like low value in the external stream. No truly sharp interface exists however, and 
consequently no corresponding surface tension. In this context the finger-like 
structures observed  in the corresponding panels of figure~\ref{fig:inst} are still 
due to the stretching of the inner core by the faster external stream. However their 
persistence cannot be ascribed to phase separation, like in the liquid-gas system.
Rather they are due to a relatively poor diffusivity of the high density features in the 
background low density environment. Indeed the essential phenomenology is 
associated with thermal diffusivity which, at fixed injection pressure, tends to smear 
out the temperature difference existing between low temperature fluid structures originated 
in the core  and the high temperature surroundings. If the process occurs too slowly 
with respect to the typical axial velocity, the density structures tend to persist for a 
significant length beyond the inlet section. This jet dynamics at supercritical conditions
is known in literature as jet disintegration~\cite{royjolseg,segpol}.

The concept is better illustrated by manipulating the continuity 
equation~\eqref{eq:cont} by inserting the expression for the  velocity divergence 
provided by  the energy equation~\eqref{eq:ener},
\begin{align}
\label{eq:cont_mod}
&{\displaystyle \pd{\R}{t}} + \vec u\cdot \nabla \rho =
-  {\displaystyle \frac{\R}{Z p}}
\frac{1}{Re Pr } \div (k \grad \T) 
\left[\frac{1}{\displaystyle{1+\frac{\left(\gamma-2\right) a \R^2}
{p \gamma}+\frac{2 a b \R^3}{p \gamma}}} \right]\, .
\end{align}
On the basis of this equation, the density adapts along a fluid trajectory due to 
thermal diffusion that forces the fluid expansion toward thermodynamic 
equilibrium with the external conditions. As a matter of fact, as the core jet 
thermodynamic state approaches the critical point, the temperature difference 
between inner and outer streams keeps on reducing for given density ratio, see the 
asterisks in comparison with open squares in the right 
panel of figure~\ref{fig:P-V_diagram}. Given the injection pressure, the material 
derivative of the density on the left hand side of the equation can be expressed as
$D \rho/Dt = \left({\partial \rho}/{\partial \T}\right)_{p}  D \T/Dt$, so 
that the effective P\'eclet number for the temperature equation,
$D\T/Dt = \left(1/Pe \right) \nabla \cdot \left(k \nabla \T \right)$, is given by
\begin{align*}
\frac{1}{Pe} = 
-  \left(\frac{\partial \T}{\partial \rho}\right)_{p}{\displaystyle \frac{\R}
{Z p}} \frac{1}{Re Pr }\left[\frac{1}{\displaystyle{1+\frac{\left(\gamma-2\right)
a \R^2}{p \gamma}+\frac{2 a b \R^3}{p \gamma}}} \right] > 0 \ .
\end{align*}
The dependence of both P\'eclet number and dimensionless thermal conductivity $k$ 
on the local temperature is non-linear and does not easily allow to predict the 
effective diffusion of the jet in presence of turbulence.
It is then worthwhile extracting from the simulation  an effective turbulent 
diffusivity $D_{eff}^*$. It is constructed with a transverse diffusion length-scale -- in 
the present case the radius of the inner injection  nozzle $R^*$ (we recall that the 
asterisk denote dimensional quantities) -- and the convective  characteristic time-scale 
$T^* = z_D^* /|\vec u^*_{core}|$ corresponding to the time needed 
by a particle travelling on the axis of the jet to reach the position where mixing between 
internal and external streams is completed. In practice the mixing length $z_D^*$  is 
quantified by the distance from the jet nozzle where  the average local 
density on the axis decreases by a prescribed amount with respect to the injection 
condition, see the isoline highlighted in figure~\ref{fig:mean_den}. The turbulent 
diffusivity follows as
$$
D_{eff}^* =  \frac{{R^*}^2|\vec u^*_{core}|}{z_D^*}
$$
providing the expression for  the turbulent P\'eclet number, 
$Pe_T = R^* |\vec u^*_{core}|/D_{eff}^* = z_D^*/R^* = z_D/R$, in terms of the 
dimensionless mixing length. The ratio of the molecular, 
$Pe = \R^* |\vec u^*_{core}| R^*/(k^*/c_p^*)$, to the turbulent  P\'eclet number 
$Pe/Pe_T = \rho^* D_{eff}^*/(k^*/c_p^*)$ measures the enhanced diffusion due to turbulence.

As anticipated, in the present case, the dimensionless mixing length $z_D^*/R^*$ 
is evaluated from the average density field as the position of the 
intercept of the selected $\rho$-isoline ($\rho^* = .95 \rho^*_{core}$) with the jet axis. 
For the four cases we find $z^*_D(A) \simeq 7.4 R^*$, $z^*_D(B) \simeq 7.9 R^*$, 
$z^*_D(C) \simeq 8.8 R^*$, $z^*_D(D) \simeq 9.8 R^*$. 
Dimensional analysis provides a list of parameters upon which $z^*_D/R^*$ may depend, namely
$$
\frac{z_D^*}{R^*} = f\left(\frac{R_1^*}{R^*},  \frac{R_2^*}{R_1^*}, 
\frac{\R_{core}^*}{\R_c^*},  \frac{\R_{core}^*}{\R_{ext}^*}, 
\frac{p_{env}^*}{p_c^*}, \frac{\R_{core}^* |\vec u^*_{core}|}{\R_{ext}^* 
|\vec u^*_{ext}|}, Re, Pr\right) \, ,
$$
where the dependence on the turbulent intensity of  the incoming inner stream is 
accounted for through the Reynolds number. In the cases we address,  geometry,  
momentum ratio ${\R_{core}^* |\vec u^*_{core}|}/({\R_{ext}^* |\vec u^*_{ext}|})$, and 
Reynolds number are held fixed. From the numerical results we find a weak dependence 
on the  pressure ratio $p_{env}^*/p_c^*$, on the Prandtl number $Pr$,  and on injection density 
$\R_{core}^*/\R_c^*$ that result in an almost constant $z_D^*/R^* \simeq 7 - 8$ 
for the three high pressure cases and a little larger for the lowest pressure
$z_D^*/R^* \simeq 9.8$.

Let us now  assume that one intends to simplify the system by  modelling the 
high-pressure, real gas 
injection with the perfect gas model for fixed mass flow rate, i.e.
given $\R_{core}^* |\vec u^*_{core}|$, and for the same injector geometry.
Two alternative  procedures can be reasonably conceived, namely keeping the same 
injection density (${\R^*}_{core}^R = {\R^*}_{core}^P$, with the superscript $R$ and $P$ denoting 
the real and the perfect gas case respectively) or, alternatively, the same temperature 
(${\T^*}_{core}^R = {\T^*}_{core}^P$). Since $z^*_D/R^*$ is more or less constant, in the 
first case (same density), where $|\vec u^*_{core}|^R = |\vec u^*_{core}|^P$, the 
persistence time of the structures $T^* = z_D^*/|\vec u^*_{core}|$ will be the same 
for the real gas and for its perfect gas model. In the 
other case (same temperature), the different injection density
$\R_{core}^{*R} > \R_{core}^{*P}$ entails a different injection speed resulting in 
a longer persistence time for the real gas. We observe that, commonly, the injection 
temperatures are the experimental control parameters, leading to an increased persistence 
of the real gas coherent structures of  density with respect to the perfect gas analogue.

\begin{figure}
\centering
\includegraphics[height=4.35cm]{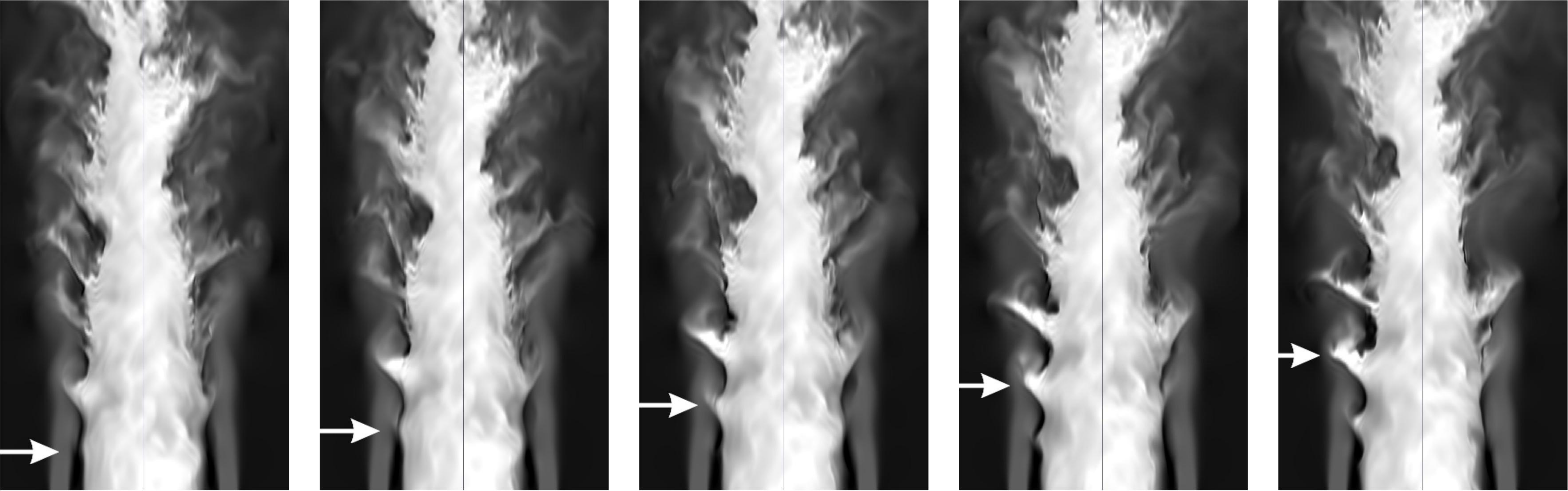}
\caption{\label{inst_evol} Real gas (Sim A) Instantaneous configurations of 
the near field of axial momentum. From left to right, consecutive instants of 
time with a time gap equal to $0.25\,R^*/|\vec{u}^*_R|$. The arrows highlight the 
evolution of a specific ``ligament''.}
\end{figure}
\begin{figure}
\centering
\includegraphics[scale=.15]{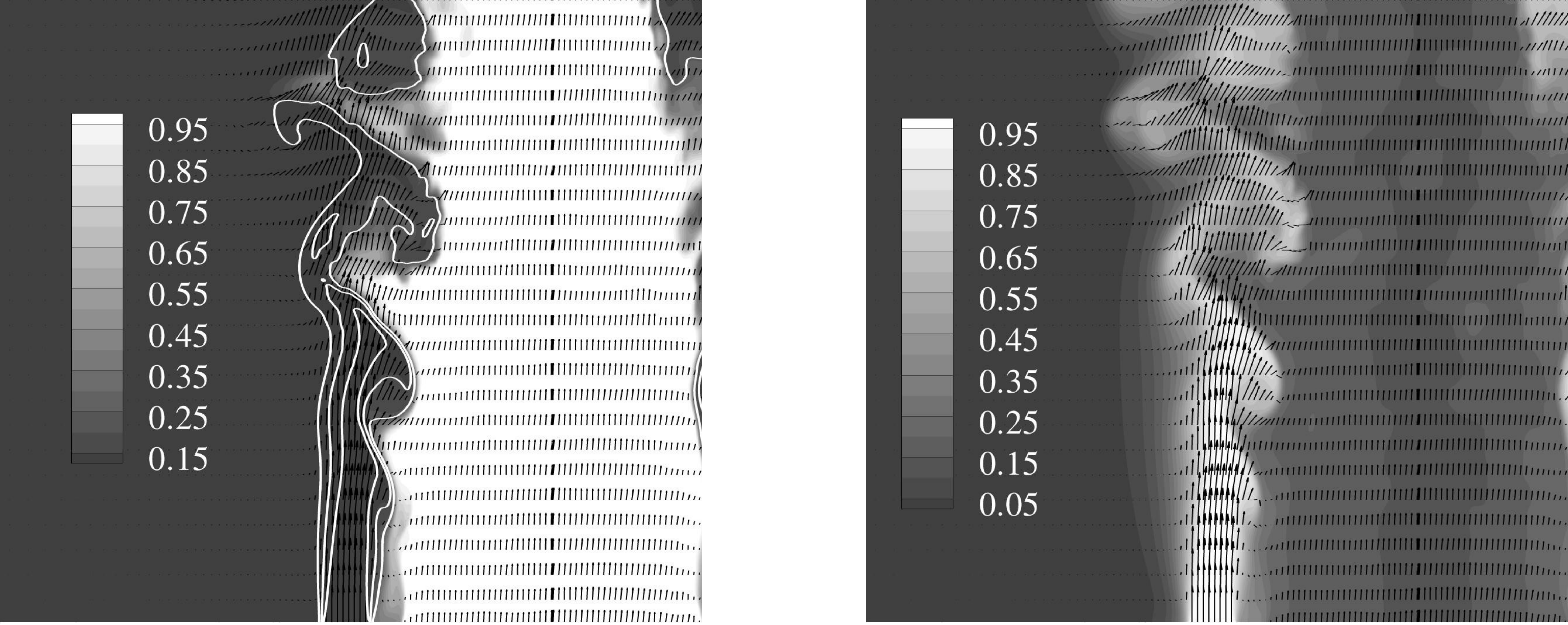}
{\scriptsize \put(-340,120){\bf (a)}}
{\scriptsize \put(-163,120){\bf (b)}}
\caption{Magnification of the region characterized by the formation of ligaments 
in the last instantaneous field shown in the rightmost panel of figure~\ref{inst_evol}. 
Panel (a): instantaneous density field with in-plane velocity vectors. 
Solid white lines denote three levels of passive scalar injected through the external 
stream. Panel (b): passive scalar contour with the in-plane velocity vectors. 
\label{fig:rhoplane}}
\end{figure}

The dynamics of ligaments formation is addressed in 
figure~\ref{inst_evol} by showing five successive configurations of
the jet near field.
Ligament formation, growth and evolution are clearly visualized by the axial momentum 
isocontours (see the arrow used to highlight the same structure in the successive  stages of 
development). The wake of the finite thickness trailing edge separating the inner from the 
outer stream, see the sketch in figure~\ref{geom_coax}, gives rise to the system of
Kelvin-Helmholtz vortices apparent in figure~\ref{fig:rhoplane}. They
force the internal, slow, and dense gas towards the external, fast, and light stream.
The extruded structures are consequently elongated by the high 
velocity stream to form the ligaments.
Panel (a) of figure~\ref{fig:rhoplane} provides the instantaneous 
density field on an axial plane, together with the in-plane velocity vectors and three 
isolevels of a passive scalar injected in the external stream. The contour plot of the passive 
scalar field is superimposed to the in-plane velocity vectors in the panel (b). 
The ligament formation is apparently correlated with the Kelvin-Helmholtz 
vortices, highlighted by the velocity vectors and by the passive scalar 
isolines. The dense structures protruding from the inner core contribute to 
slow down the external stream thereby blocking the flow and inducing additional 
radial motion.
It is worthwhile stressing that the dynamics of ligament formation is generic as long as a 
strong density and velocity contrast exists
between the two streams. Indeed, although figure~\ref{fig:rhoplane} concerns the real gas 
simulation A, a substantially identical phenomenology is found in all
the other two cases with large inner/outer density ratio (B, C). In the fourth case, D, 
persistent ligaments are not observed and density structures are much less neatly defined 
due to the small density contrast between the streams.

The most significant difference among the three high density ratio cases is found in the 
small scale features of the jet, that turn out to  be deeply influenced by the thermodynamic 
behavior of the fluid.
\begin{figure}
\centering
\includegraphics[width=.7\textwidth]{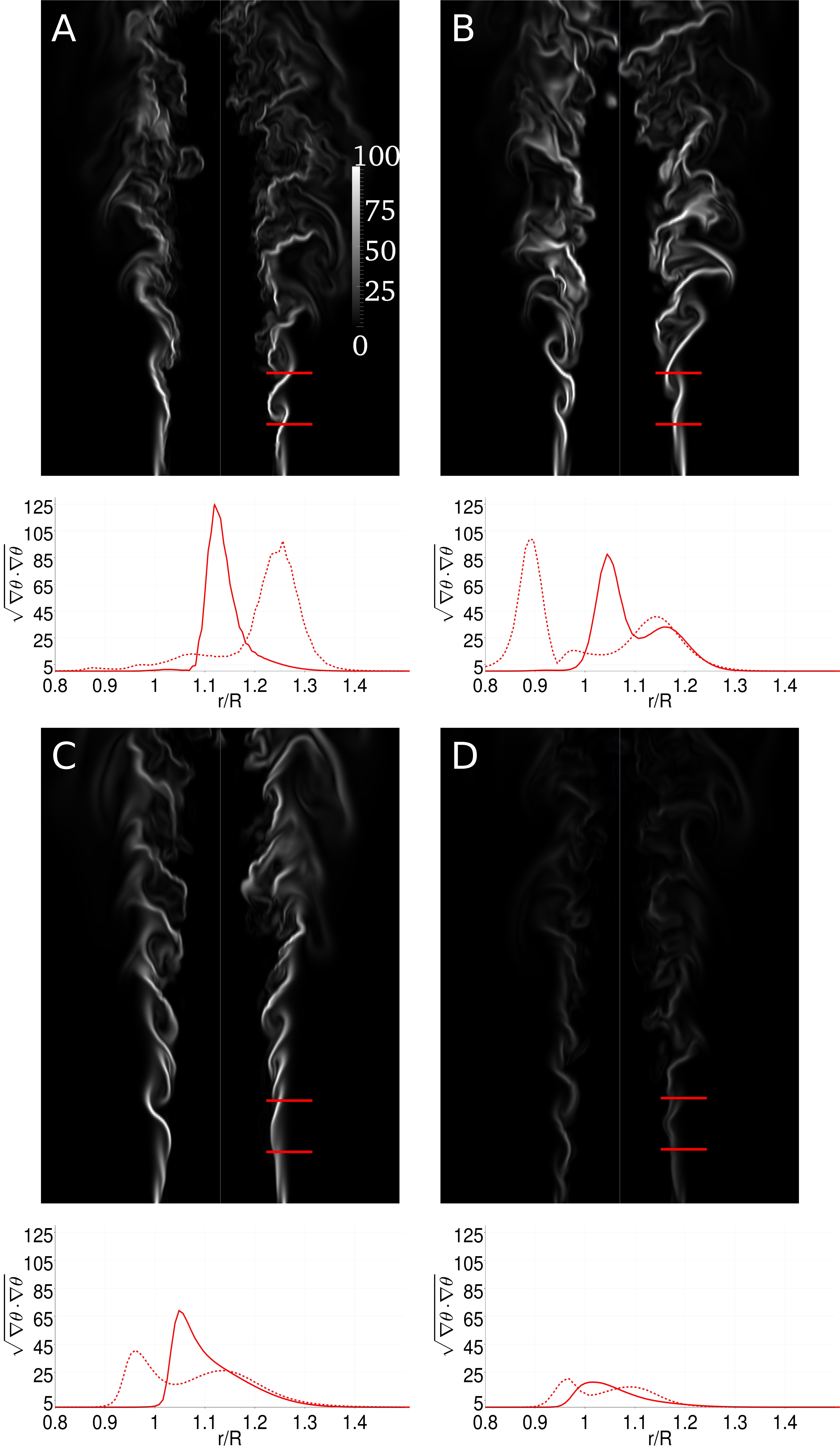}
\caption{\label{fig:gradT} Temperature gradient magnitude
$\sqrt{\vec \nabla \theta\cdot \vec\nabla \theta}$ field. From left to right and 
from top to bottom: real gas jet (Sim A), perfect gas jet (Sim B), real gas jet 
with the transport properties of the perfect gas case (Sim C), perfect gas jet 
matching the temperature ratio of the real gas case (Sim D).
The bottom part of each panel report the radial profiles of the 
temperature gradient magnitude at z=R (solid lines) and z=2R (dashed lines), in the
field contour the two axial stations are highlighted with the red segments.
}
\end{figure}
Figure~\ref{fig:gradT} addresses the magnitude of the instantaneous temperature gradients,
$\sqrt{\vec \nabla \T\cdot \vec\nabla \T}$. A peculiar aspect is that the temperature 
gradients have substantially the same order of magnitude for the two real gas cases 
(top and bottom panel in the left column) and for the perfect gas case (top right panel),
despite the temperature difference between inner and outer jet is smaller for 
the former two cases than it is for the  perfect  gas case. 
It follows that the scales at which the temperature  gradients occur tend 
to be smaller in supercritical conditions.
In order to make the argument clear, let us consider a dimensional estimate
for the temperature gradients,
$\sqrt{\vec \nabla \T\cdot \vec\nabla \T} \sim \Delta \Theta/\ell_\theta$, 
where $\Delta \Theta$ is the temperature jump across the interface
and $\ell_\theta$ is the effective thickness of the instantaneous thermal 
interface between the high and low temperature regions. Our data show that 
$\Delta \Theta^A/\ell_\theta^A \simeq  \Delta \Theta^B/\ell_\theta^B$. 
Since $\Delta \Theta^A < \Delta \Theta^B$, it follows 
$\ell_\theta^A <  \ell_\theta^B$, i.e.\ the thermal thickness for the real 
gas, supercritical jet is smaller than for the perfect gas case. 
The statistical characterization of this behavior will be provided in the following section.

\begin{figure}
\centering
\includegraphics[width=.8\textwidth]{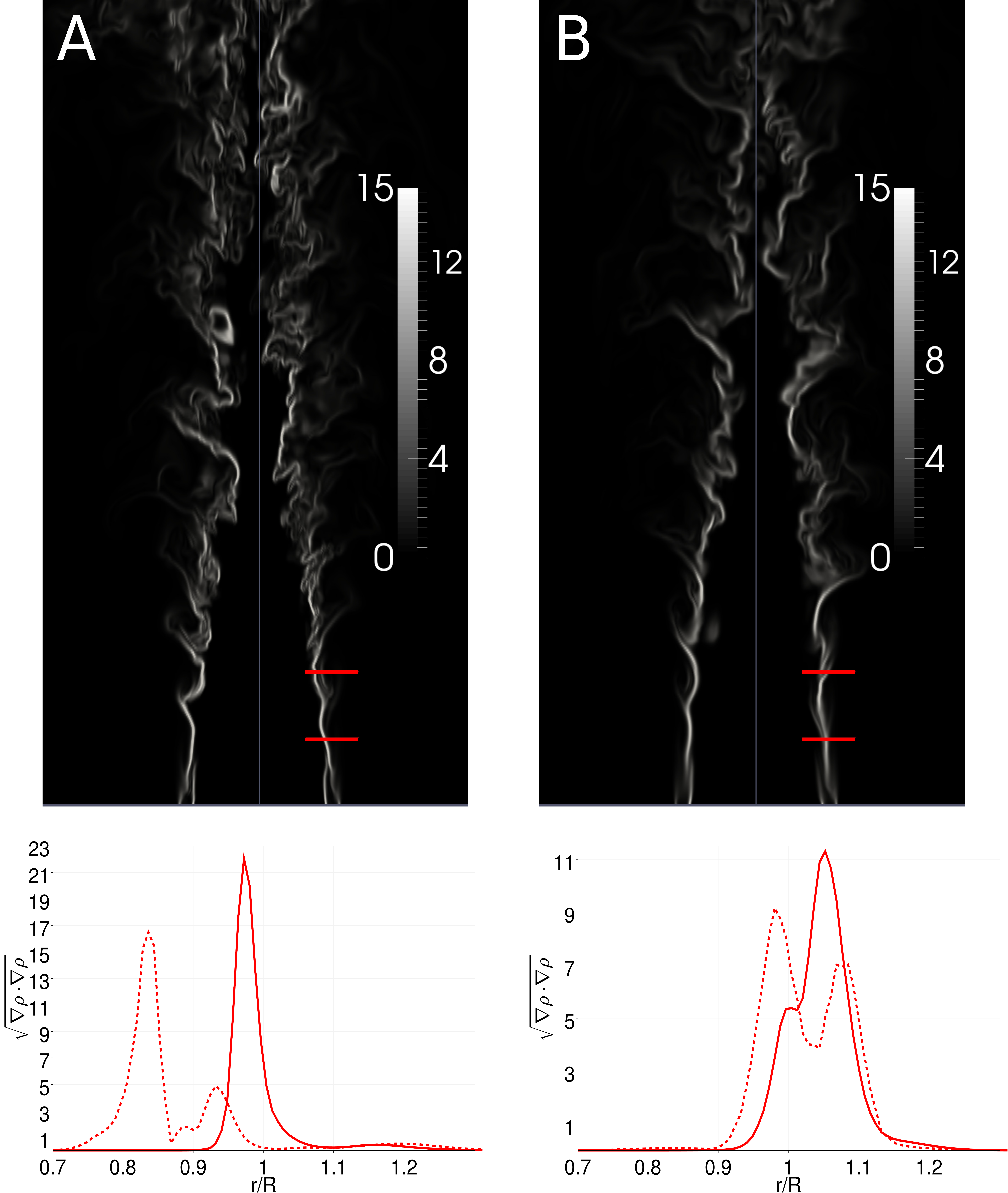}
\caption{\label{fig:density_field} Instantaneous density gradient fields of real gas jet 
(Sim A) (left top panel) and perfect gas jet (Sim B) (right top panel). The field of simulation C
is similar to the simulation A while the density gradients of simulation D are negligible respect to the
showed one, since they are omitted. In the bottom panels the radial profiles 
of the density gradient magnitude at z=R (solid lines) and z=2R (dashed lines) 
are reported, In the top panels
the two axial stations are highlighted with the red segments.}
\end{figure}

According to eq.~\eqref{eq:state} where $p(t) = const$, the density gradients 
are associated with the corresponding temperature gradients,  
$\nabla \R = \left(\partial \R(\T,p)/\partial \T\right) \nabla \T$. Here the 
real gas thermodynamics makes the difference and the singularity near the 
critical point plays a role, since $\partial \R(\T,p_c)/\partial \T \rightarrow \infty$ 
for $\T \rightarrow \T_c$, where we recall that $p_c$ and $\T_c$ are the critical 
pressure and temperature, respectively. This leads to a further intensification of 
the density gradients for the real gas near the critical point, with respect to 
the perfect gas case. Indeed, the dimensional estimate for the density thickness of 
the interface, $\ell_\R \sim  \Delta P/\sqrt{\vec \nabla \R\cdot \vec\nabla \R}$, 
where $\Delta P$ is the density jump across the interface, is 
$\ell_\R \sim \ell_\T  (\Delta P/\Delta \Theta)/\left(\partial \R/\partial \T \right)$ 
(here the symbol $P$ reads capital $\rho$). This expression shows that the 
density thickness becomes much thinner than the thermal thickness  where the gas 
gets close to the critical conditions. This behavior of the density gradients can 
be appreciated in figure~\ref{fig:density_field} that provides the isolevels of the 
gradient intensity for two instantaneous configurations, corresponding to the real 
gas and the perfect gas simulation, left and right panel, respectively. Close to 
the injection section, the density interface is apparently much sharper for the 
configuration reported on the left. Increasing the distance from the exit,
the persistent filamentary structures observed for this case become less evident 
and weaker in the perfect gas case. Meanwhile, the mixing region becomes 
increasingly convoluted by small scale, sharply contrasted details which are 
missing instead at corresponding positions in the right panel. This behavior 
corresponds to increased turbulent mixing with respect to the perfect gas case.

\subsection{Statistical analysis}

\begin{figure}
\centering
\includegraphics[width=.8\textwidth]{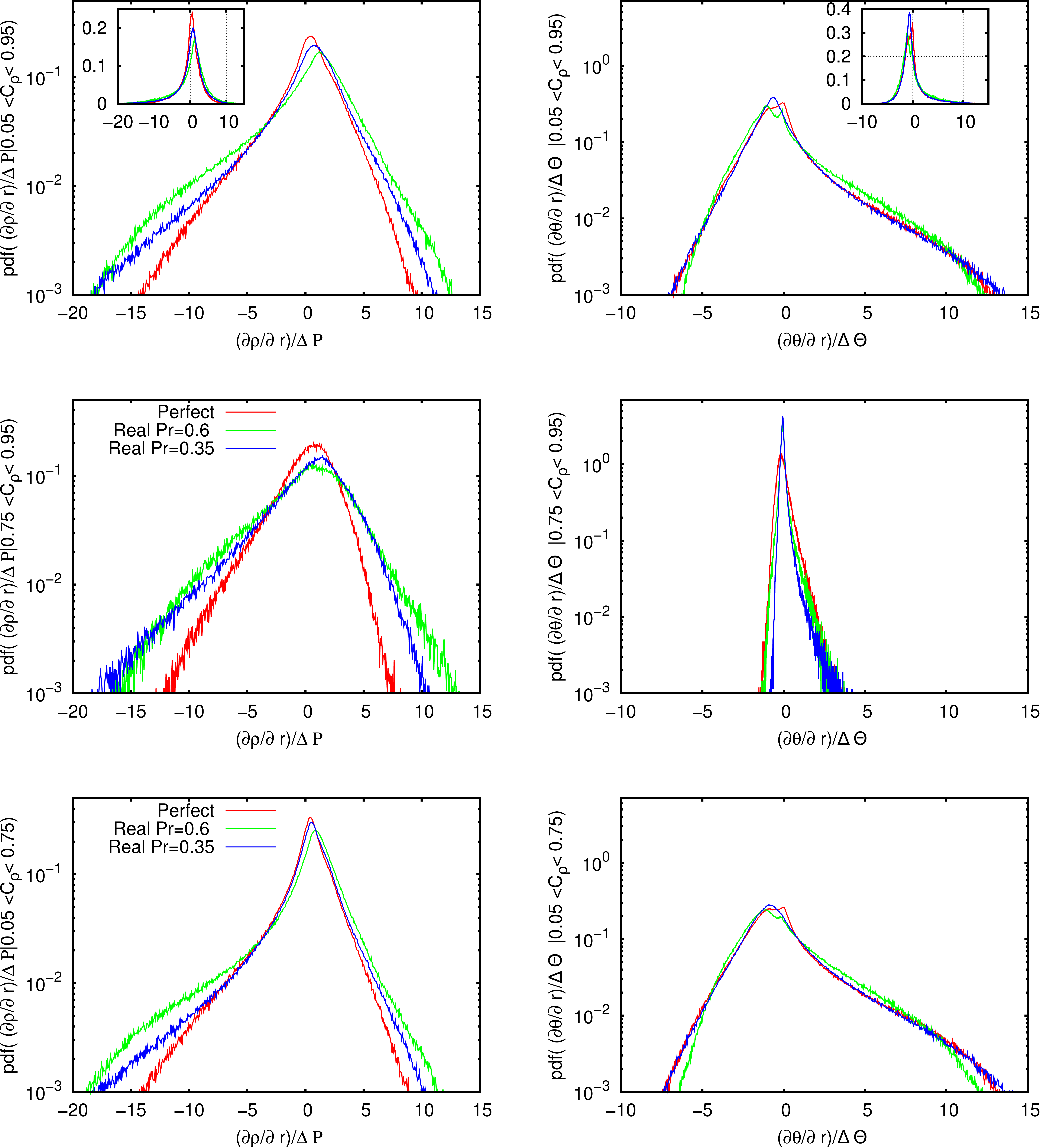}
{\scriptsize \put(-222,400){\bf (a)}}
{\scriptsize \put(-140,400){\bf (b)}}
{\scriptsize \put(-222,255){\bf (c)}}
{\scriptsize \put(-140,255){\bf (d)}}
{\scriptsize \put(-222,110){\bf (e)}}
{\scriptsize \put(-140,110){\bf (f)}}
\caption{\label{fig:pdfgrad} Probability density function 
of  the radial component of the density gradient fluctuation $\partial \R'/\partial r/\Delta P$ (left column, panels a,c,e),
and of the temperature gradient fluctuation $\partial \T'/\partial r/\Delta \Theta$ (right column, panels b,d,f).
$\Delta P=\R_{core}-\R_{ext}$ and $\Delta\Theta=\T_{core}-\T_{ext}$ are respectively the density and 
temperature difference between jet core and external environment.
 The statistics concerns the region 
extending two diameters downstream the inlet section $0<z<4R$ and is conditioned to different 
ranges of the density, namely $0.05<C_\R<0.95$ (top; panels a,b), $0.75<C_\R<0.95$ (middle; panels c,d) and $0.05<C_\R<0.75$ 
(bottom; panels e,f), where $C_\R = \left(\R-\R_{ext}\right)/\Delta P$ is the normalized density  ranging from $C_\R=0$  
in the external environment to $C_\R=1$ in the jet core. 
}
\end{figure}

The previous subsection dealt  with the instantaneous configurations of the jets. 
The conclusions we reached are now better substantiated by quantitatively addressing 
the related statistics. The statistical analysis is based on the same collection of about two hundred
instantaneous fields, separated in time  by $0.25\,R^*/|\vec{u}^*_R|$, used for the mean fields.

The panels in the left column of Figure~\ref{fig:pdfgrad} show the probability 
density function of the radial component of the normalized density fluctuation gradient, 
$(\partial \R'/\partial r)/\Delta P$, with $\R' = \R - \langle \R\rangle$ and 
$\langle \R \rangle$ the local mean density. As a matter of fact the density gradient is, on 
average, almost aligned with the radial direction suggesting that the statistics of $\partial 
\R'/\partial r$ conveys most of the information on the structure of the 
instantaneous field. The statistics concerns the region extending two diameters 
downstream the inlet section ($0 \le z \le 4 R$) and is conditioned to different 
density ranges, namely $0.05<C_\R<0.95$ (top panel), $0.75<C_\R<0.95$ 
(middle panel), $0.05<C_\R<0.75$ (bottom panel), 
with $C_\R = \left(\R- \R_{ext} \right)/\Delta P$ the density normalized such that $C_\R = 0$ in the external 
stream and  $C_\R = 1$ in the core. 
The panels in the right column of Figure~\ref{fig:pdfgrad}  refer to the 
radial component of the temperature fluctuation gradient, $\partial \T'/\partial r/\Delta \Theta$, with 
$\T'=\T-\langle\T\rangle$ the temperature fluctuation with respect to the mean one $\langle\T\rangle$
and $\Delta \Theta=\T_{core}-\T_{ext}$.
Conditioning to density is indeed instrumental to focus the analysis on the instantaneous interface 
between high density inner core  and low density outer stream.
This is the region of the phase space where well defined ligaments tend to be formed, see 
Figure~\ref{fig:density_field}.
In the three cases (A, real gas; B, perfect gas; C, real gas with artificial transport properties) the pdf 
$f(\partial \rho'/\partial r/\Delta P | 0.05 \le C_\R \le 0.95, 0 \le z \le 4 R)$ is significantly skewed towards the 
negative tail, see top panel on the left and the related inset showing the pdf in  logarithmic and linear 
scale, respectively.
The real gas cases show longer tails indicating  large intensity events comparatively more frequent than in the perfect gas case.
This behavior is quantified by the normalized even moments of the pdf,  the so-called hyper-flatness factors of order $n$
\begin{align}
F_{2n}[\partial \rho'/\partial r] = \frac{\langle \left(\partial \rho'/\partial r\right)^{2n} \rangle}{\langle \left( \partial \rho'/\partial r\right)^{2} 
\rangle^n} \, \qquad \qquad
F_{2n}[\partial \theta'/\partial r] = \frac{\langle \left(\partial \theta'/\partial r\right)^{2n} \rangle}{\langle \left( \partial \theta'/\partial r\right)^2\rangle^{n} } \, ,
\end{align}
see the data reported in Table~\ref{tab:flatness}. Values of the hyper-flatness 
exceeding the reference values for a Gaussian distribution, 
$F_{2n}^G = 3, 15, 105$ for $n=2,3,4$,  signal the existence of an intermittent 
behavior, where phases of relatively weak gradients are alternated with the presence 
of relatively rare but intense events. In our system the origin of the intermittency 
is related to the ligaments that invade regions of relatively smooth density variation.
\begin{table}
\centering
\begin{tabular}{c|rrr}
           &n=2&n=3&n=4\\\hline
sim. A&7.2&110.&2327.\\
sim. B&6.4&  91.&1930.\\
sim. C&5.1&  49.&  624.\\
Gauss&3.0& 15.&105.\\
\end{tabular}\hspace{1.cm}
\begin{tabular}{c|rrr}
           &n=2&n=3&n=4\\\hline
sim. A&9.0&172.&4464.\\
sim. B&8.1&134.&2990.\\
sim. C&6.0&72.&1151.\\
Gauss&3.0&15.&105.\\
\end{tabular}
\caption{\label{tab:flatness} Hyper-flatness, $F_{2n}\left[\partial \cdot'/\partial r\right]$,
 of the probability distribution reported in the top panels of figure~\ref{fig:pdfgrad}. 
Left table: radial component of density gradient 
fluctuations; right table: radial component of temperature gradient fluctuations.}
\end{table}

Given the relationship between temperature and density gradients, 
$\partial \T'/\partial r = \left(\partial \T/\partial \R\right) \partial \R'/\partial r$, the overall shape of 
the temperature pdf is grossly speaking specular with respect to that of density since  
$\partial \T/\partial \R < 0$. Interestingly, the three temperature gradient pdfs for the two real gas 
and the perfect gas case are substantially identical, confirming the impression gained from inspecting 
the instantaneous field, see Figure~\ref{fig:gradT}.
We stress that  conditioning with respect to density is used  in the two  top panels at the only purpose of 
removing from the analysis the events of vanishing gradients occurring in the external region and in the 
inner jet core that would otherwise outnumber the comparatively less numerous events belonging to the 
physically significant interface between inner and outer stream.

In order to focus on the large density features, the middle panels of Figure~\ref{fig:pdfgrad} show 
the fluctuation gradient pdf conditioned to the high density range $0.75<C_\R<0.95$. It is apparent that 
the origin of the intermittent behavior of the  density gradients (left panel) is mostly related to features 
associated  with large density.  The pdf shows exponential tails which are definitely longer in the two real 
gas cases. Conversely, the temperature gradients conditioned to the same density range (right panel) are 
characterized by a comparatively narrower distribution,  indicating that the structures which support the 
density gradients have almost deterministic temperature gradients,
$\partial \T/\partial r \simeq \partial \langle \T \rangle/\partial r$, as follows from observing that $\partial \theta'/\partial r \simeq 0$.
Considered that  
$\partial \langle \T \rangle/\partial r = \langle\partial \T /\partial  \R \, \partial \R/ \partial r \rangle$, since 
$\partial \R /\partial  \T$ is large for the near critical gas, it follows that the high density features 
supporting the density gradients  are almost isothermal in the real gas cases.

The bottom panels of Figure~\ref{fig:pdfgrad} concern the complementary density range  
$0.05<C_\R<0.75$. Apparently the most significant fluctuations of the temperature gradient (left panel)  
occur in this region of the phase space. The intermittency of the temperature gradient is 
significant, implying that, locally, a large temperature difference occurs, with relatively low 
temperature regions getting close to high temperature ones. Clearly, this induces strong density gradients, 
giving reason of the non-negligible density-gradient intermittency found also in this complementary 
density range.

The density gradient is strictly related to the thickness of the interface between the high-density
core and the low-density external stream. Indeed the high density gradients that characterize
the real gas jets suggest that the interface is thinner for the real gas than for  the perfect gas.
The interface thickness  can be estimated  as the inverse of the normalized density gradient 
module $\delta_\R=\Delta P/\left|\nabla \R\right|$.
\begin{figure}
\centering
\includegraphics[width=.925\textwidth]{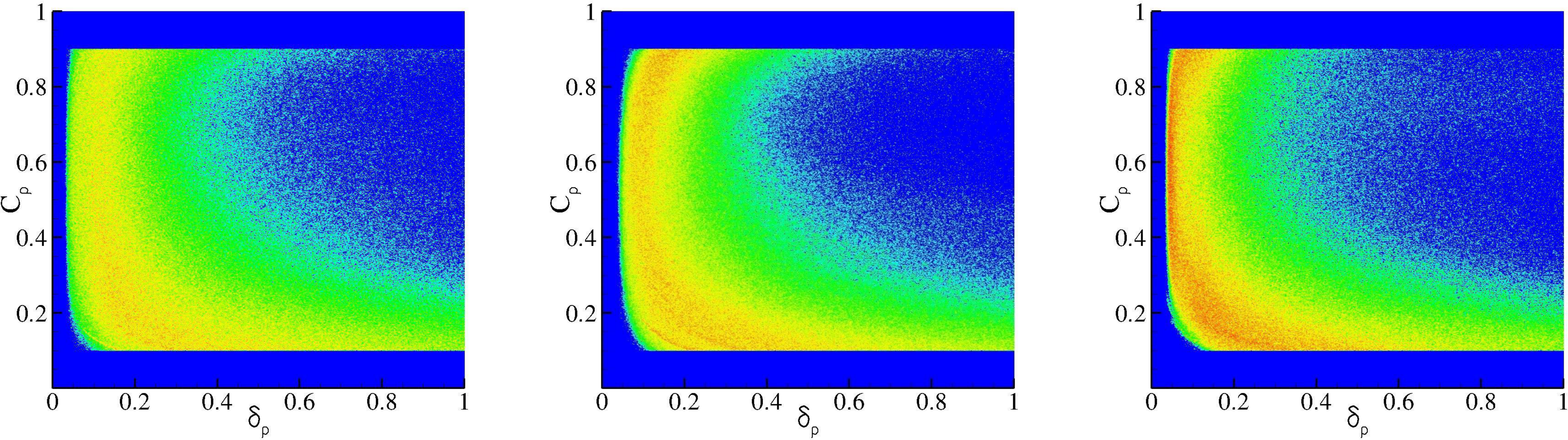}
{\scriptsize \put(-445,110){\bf (a)}}
{\scriptsize \put(-295,110){\bf (b)}}
{\scriptsize \put(-140,110){\bf (c)}}
\caption{\label{fig:pdf_width} Joint probability density function, 
$f(\delta_\rho, C_\rho)$, of  density interface thickness 
$\delta_\R=\Delta P/\left|\nabla \R\right|$ and normalized density $C_\R$, in 
the region one diameter downstream the jet inlet. From left to right: real gas 
jet (panel a), perfect gas jet (panel b) and real gas jet with perfect gas 
transport features (panel c).
}
\end{figure}
Figure \ref{fig:pdf_width} provides the joint probability density function, 
$f(\delta_\rho, C_\rho)$, of interface thickness $\delta_\R$ and normalized 
density $C_\R$, where the real gas case A is reported on the panel (a), 
the perfect gas case B in the panel (b), and the real gas case with perfect 
gas transport coefficients C in the panel (c). 
The statistics concern the same region of the 
flow domain already considered in the pdfs of figure~\ref{fig:pdfgrad} and 
are conditioned to a density interval ranging from $C_\R=0.9$ to $C_\R=0.1$.

In order to understand the different roles played by the equation of state and 
by the transport coefficients it is worth first comparing real and perfect gas 
cases with identical transport coefficients, cases  B and C reported in the 
middle and right-most panel of the figure, respectively. Apparently as the 
density increases ($C\R$), the interface thickness decreases for both cases. 
In comparison with the perfect gas case, at given normalized density, the pdf 
of the interface thickness is much more peaked at small values for the reals 
gas case. A further feature to be noted is that, at large density, the peak of 
the pdf moves toward slightly larger scales for the perfect gas case, implying 
an increased diffusion of high density features. On the contrary the pdf peak 
remains centered on the smallest scales for the real gas. This behavior is 
explained by looking at the  isobars in the density-temperature diagram 
provided in figure~\ref{fig:P-V_diagram}. The injection states for cases B and 
C are denoted with the squares and the asterisks, respectively, reported on the 
same isobar (green curve). Apparently, huge density difference take place at 
almost constant temperature for case C (real gas), see also the inset with the 
isobars plotted in linear scale. On the contrary for case B density changes are 
almost uniformly distributed along the temperature range. Since diffusion effects 
are determined by the temperature field, eq.~(\ref{eq:cont_mod}), it follows that 
density structures are more diffusive for case B, where they are always associated 
with significant temperature differences. The comparatively smaller diffusion taking 
place in the real gas explains the statistically smaller interface thickness observed 
in the scatter plots shown in the panel (c) of figure~\ref{fig:pdf_width} in 
comparison with that shown in the panel (b). The effect is observed at all 
values of density a part from the smallest range farther from the critical condition, 
where the isobar gets closer to the behavior of the perfect gas. 

When recovering the actual transport coefficients of the real gas (case A), 
panel (a) of figure~\ref{fig:pdf_width}, the typical scales of the density gradients
increase significantly with respect to the artificial case C (real gas with perfect 
gas transport coefficients). The effect is qualitatively explained by comparing 
the behavior of the thermal diffusivity $\lambda$ for the real gas shown in the 
right panel of figure~\ref{fig:visc_ther}, see Appendix~\ref{sec:visc_ther}, with 
the Sutherland law shown as a dashed line in the same figure. Clearly the diffusivity 
is enhanced, hence the increased scale for the density gradients.

In conclusion, the coherent structures of high density observed as a consequence 
of the high density and momentum contrast between inner and outer stream are 
characterized by a sharp interface of separation with the background low density 
environment. The instantaneous interface is extremely sharp for the real gas 
(Van der Waals equation of state), although it is partially smeared by the increased 
thermal diffusivity occurring near the critical point. Such features of coherent 
density and their sharp interface are responsible for the highly intermittent 
statistical behavior observed for the density gradients.

\section{Final Remarks}
\label{sec:conclusions}

The low Mach number asymptotic expansion of the Navier-Stokes equations, originally derived by Majda \&
Sethian for perfect gas flows in reactive conditions, was here extended to a generic real gas equation 
of state to deal with fluids in near critical conditions.
The resulting formulation allowed the detailed analysis of the turbulent mixing of slightly supercritical fluids at vanishing Mach number
through accurate and efficient Direct Numerical Simulations of a coaxial jet of a single 
component Van der Waals fluid. When slightly supercritical streams are mixed at large Reynolds number, turbulence
and real-gas effects combine and have been shown to produce peculiar effects. 

Elongated 
finger-like structures, the so-called ``ligaments'', similar to those classically 
observed in the break-up of liquid jets, are found in the simulations.
These structures have been well documented also in supercritical injection experiments 
and numerical simulation at moderately high Mach number and differ from those characterizing  
liquid jets which eventually break up in droplets. The supercritical ligaments, 
once formed, diffuse to eventually disappear altogether. The ligaments are originated 
by the joint effect of a high density contrast between inner and outer stream in 
combination with the strong shear layer generated at the interface between the slower inner 
jet and the outer faster stream. The high velocity ratio of the shear layer promotes the 
formation of the classical Kelvin-Helmholtz instability with rolling vortices. Finger-shaped 
high density protrusions are extruded by these vortices in the low density high speed 
stream where they are elongated well inside the external environment thereby generating the 
ligaments. It is worth emphasizing that  this mechanism operates in all the cases where 
a high density contrast exists, irrespective of the thermodynamic model. The main difference 
between Van der Waals and perfect gas jets mostly resides in the small-scale features of the 
ligaments. In comparison with the perfect gas model, the dense gas case shows much steeper density 
gradients and a thinner interface between the high- and low-density streams. 
Instead  temperature gradients behave much more similarly in the 
two cases, despite the fact that the bulk temperature differences are smaller in the 
dense gas case, for given density contrast. 
In other words, turbulent mixing in dense gases is associated with 
smaller scales than in perfect gases in order to balance the smaller temperature jump 
occurring at near critical conditions for given density contrast between the mixing streams.
The steep density gradients near the critical point are indeed associated with the 
critical point singularity that controls the spatial scales of the density variations.

The statistical signature of this phenomenology is found in the intermittency of 
the density field, as evaluated by the hyper-flatness of the density gradient pdf. 
Indeed the increased intermittency in real gas mixing is related to the occurrence 
of the mentioned ligamentary structures characterized by an extremely sharp interface separating 
high and low density streams. The picture is reinforced after looking at the large 
density gradient fluctuations that take place in correlation with high density regions, 
thermodynamically closer to the critical state. Typically, the ligaments are almost 
isothermal structures with small temperature differences with the local environment.
This behavior is understood after considering that fluid particles at near critical 
conditions may assume significantly different densities with almost identical 
temperatures as a consequence of the aforementioned critical point singularity.
Overall, the much smaller local interface thickness  for the near critical gas may have 
significant consequences for the numerical simulation of this kind of flows.

As a final remark, we like to stress that the features that have been observed 
in connection with the Wan der Waals model for the gas are expected to hold also 
for other thermodynamic model of dense gases (e.g. the Peng-Robinson model), that 
in certain cases may provide a better description of a real gas.

The intermittent behavior promoted  by the ligaments in combination with the sharp interface separating high and low 
density regions is a crucial feature to be modeled in view of increasing the  predictive performances of coarse-grained 
descriptions like Large-Eddy-Simulation. 
The formulation here introduced could indeed represent a solid framework to develop appropriate sub-grid models to deal with
turbulent diffusion  processes in high-Reynolds-number  technological applications involving supercritical fluids.

\section*{Acknowledgements}
The authors acknowledge the CASPUR High Performance Computing Centre for the 
computational resources provided via \emph{std10-284} grant. 

\appendix
\section{Van der Waals and Peng-Robinson equation of state}
\label{sec:appendix VdW}

Thermal properties of polyatomic gases are affected by quantum effects that 
emerge already at ordinary thermodynamic conditions. In addition, when 
the density is sufficiently high, molecule-molecule interaction become significant 
giving rise to the so-called real-gas effects. All this information is gathered
in the expression for the Helmholtz free-energy as a function of temperature, 
volume and molecule number. 

\subsection{The Van der Waals model}

The simplest model endowed with all these features is a diatomic Van der Waals 
gas, whose Helmholtz free energy $f$ can be derived from quantum statistical 
mechanics considerations\cite{zhu}, 
\begin{align}
f&= {\cal R} n \T \ln{ \left[
\frac{\left(1-e^{-\frac{\tau}{\T}} \right)}{C(n) \left(V-b \right) \T^{5/2}}
\right]} 
- \frac{a}{V} + {\cal R} n\frac{\tau}{2}
\label{eq:free_hel}
\end{align}
\n where $C(n) = 8 \pi^2 n \iota {\cal R}^{5/2}  \sqrt{2\pi m}/\left(\hbar^5 
 N_A^{3/2} \right) $ with $\iota$ and $m$ the moment of inertia and the mass of 
the molecule, ${\cal R}$ the universal gas constant, $\hbar = h/(2 \pi)$ with 
$h$ the Planck's constant, $N_A$  the Avogadro's number, $n = N/N_A$ the number 
of moles, $N$ the number of gas molecules, $\tau=\hbar \zeta N_A/{\cal R}$ where 
$\zeta$ is the fundamental vibrational frequency of the molecule. In the expression 
for the free-energy two additional constant appear, $a$ and $b$ related to the 
intermolecular forces and to the excluded volume, respectively.

The pressure equation of state follows as
\begin{align}
p=-\left.\pd{f}{V}\right|_{\T,N}
=\frac{{\cal R}_m \T \R}{1-b'\R}-a'\R^{2}\, ,
\label{eq:p_rho_app}
\end{align}
where $a'=a/(n {\cal W})^2$ and $b'=b/(n {\cal W})$ with $\cal W$ the molar mass  
respectively, and ${\cal R}_m={\cal R}/{\cal W}$. The entropy is
\begin{align}
{\cal S}=-\left.\pd{f}{\T}\right|_{V,N}&=
 \frac{5}{2} {\cal R} n \left(\ln{\T}+1\right) 
+ {\cal R} n \ln{\left[ \frac {C(n)  \left(V-b\right)} {\left(1-e^{-\frac{\tau}{\T}}\right)}\right] }+
\frac{{\cal R} n\tau}{\left(e^{\tau/\T}-1\right)\T} \,,
\label{eq:entropy}
\end{align}
which yields the internal energy 
\begin{align}
U= F+\T S = \frac{5}{2}{\cal R} n \T - \frac{a}{V} +{\cal R} n \tau 
\left(\frac{1}{2}- \frac{1}{e^{\tau/\T}-1}\right) \ .
\label{eq:Int_Ene}
\end{align}
\n Hence the heat capacity at constant pressure and volume 
can be calculated from the entropy by means the known thermodynamic 
relations,
\begin{align}
\label{eq:cv}
c_v&=\T \left.\pd{\cal S}{\T}\right|_{V,N}= 
\frac{5}{2} {\cal R} n  
+{\cal R} n \frac{\tau^2}{\left(e^{\tau/\T}-1\right)^2 \T^2}\\
\label{eq:cp}
c_p&=c_v + \T \left.\pd{p}{\T}\right|_{V,N}\left.\pd{V}{\T}\right|_{p,N}
=c_v + \T \frac{{\cal R}^2 n^2}{\left(V-b\right)\left(p-a/V^2+2ba/V^3\right)} \ .
\end{align}
The ideal gas is recovered by  setting $a = b = 0$,  with $\tau = 0$ 
recovering constant, temperature-independent thermal properties.
Relations \eqr{p_rho_app}, \eqr{cv} and \eqr{cp} can be re-expressed as 
\begin{align}
\label{eq:p_rho_app_2}
p&=\frac{{\cal R}_m \T \R}{1-b'\R}-a'\R^2\, ,\\
\label{eq:cv_rho}
c_v'&= \frac{c_v}{n \cal W}= \frac{5}{2} {\cal R}_m  
+{\cal R}_m \frac{\tau^2}{\left(e^{\tau/\T}-1\right)^2 \T^2}\, ,\\
\label{eq:cp_rho}
c_p'&=\frac{c_p}{n \cal W}=c_v' + {\cal R}_m \frac{{\cal R}_m \R  \T}
{\left(1-b'\R\right)\left(p-a'\R^2+2b'a'\R^3\right)}\, ,
\end{align}
where $c_v'=c_v/(n {\cal W})$ and $c_p'=c_p/(n {\cal W})$ are the heat 
capacity coefficients per mass unit. 

\subsection{The Peng-Robison model}
\label{sec:peng_robinson}

In many circumstances the Van der Waals model is not suffuciently accurate, and should be substituted by alternative thermodynamic models.
As an example, the pressure equation for the Peng-Robinson model, \cite{penrob}, reads
\begin{align}
p=\frac{k_B N \T}{V - b}-\frac{a \alpha(\T)}{V+2\,b\,V-b^2}\, ,
\label{eq:pg}
\end{align}
where $V$ is the volume and $a$, $b$ and $\alpha(\T)$ can be expressed 
as a function of the critical thermodynamic variables and of what is 
called the acentric factor $\omega$,
\begin{align*}
a &=\frac{0.457235 N^2 k_B^2 \T_c^2}{p_c}\\
b &=\frac{0.077796 N k_B \T_c}{p_c}\\
\alpha&= \left[1+\kappa\left(1-\T_R^{0.5}\right)\right]^2  \ ,
\end{align*}
with $\T_R={\T}/{\T_c}$ and $\kappa=0.37464+1.54226\,\omega-0.26992\,
\omega^2$. Considering the relation  $p = - (\partial f/\partial V)_{\T,N}$, 
the Helmholtz free-energy $f$ associated with the equation of state 
\eqref{eq:pg} is obtained by straightforward integration,
\begin{align}
f = -k_B \theta \ln{{\cal Z}}=- k_B \theta N\,\int{\frac{1}{V-b}dV}+
a\,\alpha(\T)\int{\frac{1}{V+2\,b\,V-b^2}dV}+ A(\T,N) \, ,
\end{align}
where $A(\T,N)$ is the integration constant which depends only on temperature 
and particle number. The result is
\begin{align}
f= - k_B \T N \ln{\left|V-b\right|}+\frac{a\alpha(\T)}{2\sqrt{2}\, b}
\ln{\left|\frac{V-\left(\sqrt{2}-1\right)\,b}{V+\left(\sqrt{2}+1\right)\,b}\right|}+A(\T,N)\,.
\label{eq:z_pr}
\end{align}
The integration constant  $A(\T,N)$ may be evaluated considering that in the limit 
$V\rightarrow\infty$ for fixed temperature and particle number both the 
Peng-Robinson and the Van der Waals models should approach the same limit. 
In other words, in the limit,
equations \eqr{free_hel} and \eqr{z_pr} must eventually coincide, providing by comparison the expression for $A$
\begin{align*}
&A(\T,N)=- \frac{5}{2} k_B \T N\ln{\T} + k_B N \frac{\tau}{2} + k_B \T N\,\ln{\left(1-e^{-\frac{\tau}
{\T}}\right)} - k_B \T  N \ln{\left(\frac{8}{\hbar^5} \pi^2 N \iota k_B^{5/2} 
\sqrt{2\pi m}\right)}\,.
\end{align*}
Once $A$ is found, all the relevant thermodynamic quantities  are accessible from the Helmholtz free-energy that reads
\begin{align}
\notag
f=&{\cal R} n \T  \ln{\left|V-b\right|}-
\frac{a\alpha(\T)}{\,2\sqrt{2}\, b}
\ln{\left|\frac{V-\left(\sqrt{2}-1\right)\,b}
{V+\left(\sqrt{2}+1\right)\,b}\right|}\\
+& \frac{5}{2} {\cal R} n \T \ln{\T} - 
 {\cal R} n  \frac{\tau}{2} - 
 {\cal R} n \T \,\ln{\left(1-e^{-\frac{\tau}{\T}}\right)} 
+{\cal R} n \T   \ln{\left(\frac{8}{\hbar^5} \pi^2 N \iota k_B^{5/2}
\sqrt{2\pi m}\right)}\, .
\label{eq:hel_Peng}
\end{align}

\section{Transport coefficients.}
\label{sec:visc_ther}

The dynamic viscosity and thermal conductivity are evaluated, see Ref~\onlinecite{lemjac}, 
considering the dilute gas contribution, $\mu^0$ and $k^0$, the 
residual fluid contribution, $\mu^r$ and $k^r$, and the critical state 
contribution, $k^c$. The critical state enhancement can be neglected 
for the dynamic viscosity.

For the viscosity the dilute gas contribution is given by
\begin{align}
\mu^0(\T) = \frac{0.0266958\sqrt{{\cal W}\T}}{\sigma^2 \Omega(\bar{\T})}\, ,
\label{eq:eta0}
\end{align}
where $\sigma$ is the Lennard-Jones size parameter equal to 
$\sigma=0.3656\,{\rm nm}$ for the Nitrogen.
$\Omega$ is the collisional integral 
\begin{align}
\Omega(\bar{\T})=\exp{\left(\sum_{i=0}^4 b_i \left[\ln(\bar{\T})\right]^i\right)}\, ,
\label{eq:omega_eta}
\end{align}
where $\bar{\T}=\T/(\varepsilon/k)$ and $\varepsilon/k$ is the Lennard-Jones energy parameter 
equal to $\varepsilon/k=98.94 K$ for the Nitrogen. $b_i$ are fitting coefficient provided 
in Ref~\onlinecite{lemjac}, and are reported in table~\ref{tab:bi}.
\begin{table}
\centering
\begin{tabular}{c|c}
$i$ & $b_i$ \\ \hline
$0$ & $0.431$ \\ 
$1$ & $-0.4623$ \\ 
$2$ & $0.08406$ \\ 
$3$ & $0.005341$ \\ 
$4$ & $-0.00331$ 
\end{tabular}
\caption{\label{tab:bi} Fitting coefficients of the Collision integral for the dilute gas 
viscosity provided in~\cite{lemjac}.}
\end{table}
The residual contribution to the dynamic viscosity yields
\begin{align}
\mu^r(\tau,\delta)=\sum_{i=1}^n N_i \tau^{t_i}\delta^{d_i} 
\exp{\left(-\gamma_i\delta^{l_i}\right)}\, ,
\label{eq:etar}
\end{align}
where $\tau$ and $\delta$ are the ratios $\tau=\T_c/\T$ and $\delta=\R/\R_c$, respectively, while
the coefficients $N_i$, $t_i$, $d_i$, $l_i$ and $\gamma_i$ are reported in table~\ref{tab:coeff}.
\begin{table}
\centering
\begin{tabular}{c|ccccc}
$i$ & $N_i$        & $t_i$  & $d_i$ & $l_i$ & $\gamma_i$\\ \hline
$1$ & $10.72$      & $0.1$  & $2$   & $0$   & $0$ \\ 
$2$ & $0.03989$    & $0.25$ & $10$  & $1$   & $1$ \\ 
$3$ & $0.001208$   & $3.2$  & $12$  & $1$   & $1$ \\ 
$4$ & $-7.402$     & $0.9$  & $2$   & $2$   & $1$ \\ 
$5$ & $4.620$      & $0.3$  & $1$   & $3$   & $1$   
\end{tabular}
\caption{\label{tab:coeff} Coefficients and exponent of the residual viscosity equation.}
\end{table}
The dynamic viscosity is obtained by the sum of the two contribution obtaning the viscosity in 
$\mu Pa\cdot s$. The model for the thermal conductivity, espressed in $mW/(m\cdot K)$, is composed of
three contribution. The first one is the dilute gas contribution,
\begin{align}
k^0=N_1\left[\frac{\mu^0(\T)}{1 \mu Pa\cdot s}\right]+N_2 \tau^{t_2} +N_3 \tau^{t_3}\, .
\label{eq:lambda}
\end{align}
The second contribution to the thermal conductivity equation is the residual one,
\begin{align}
k^r=\sum_{i=4}^n N_i \tau^{t_i} \delta_{d_i} exp{\left(-\gamma_i\delta^{l_i}\right)}\,.
\label{eq:lambdar}
\end{align}
The third contribution dealing with the critical state correction reads
\begin{align}
k^c=\R c_p \frac{k_B R_0 \T}{6 \pi \xi \mu(\T,\R)}\left(\tilde{\Omega}-\tilde{\Omega}_0\right)
\label{eq:lambdac}
\end{align}
where
\begin{align}
\label{eq:omega_ther}
\tilde{\Omega}  &=\frac{2}{\pi}\left[\left(\frac{c_p-c_v}{c_p}\right)\tan^{-1}\left({\frac{\xi}{q_D}}\right)+\left(\frac{c_v}{c_p}\right)\left(\frac{\xi}{q_D}\right)\right]\\ 
\notag&\\
\label{eq:omega0_ther}
\tilde{\Omega}_0&=\frac{2}{\pi}\left\{1-\exp{\left[\frac{-1}{(\xi/q_D)^{-1}+(\xi/q_D)^{2}/3(\R_c/\R)^2}\right]}\right\}
\end{align}
\begin{align}
\xi=\xi_0\left[\frac{\tilde{\chi}(\T,\R)-\tilde{\chi}(\T_{ref},\rho)\T_{ref}/\T}{\Gamma}\right]^{\nu/\gamma}
\label{eq:xi}
\end{align}
\begin{align}
\tilde{\chi}(\T,\R)=\frac{p_c \R}{\R_c^2} \left.\pd{\R}{p}\right|_\T \ .
\label{eq:chi}
\end{align}
The coefficient and exponents of these equations are summarized in table~\ref{tab:coeff_ther}, while the other relevant parameters
are the Boltzmann's constant $k_B=1.380658 \cdot 10^{-23} J/K$, the constants 
$R_0=1.01$, $\nu=0.63$ and $\gamma=1.2415$.  Finally a few other constants  depend on the specific gas and data fitting on Nitrogen yields $q_D=0.4 nm$, $\xi_0=0.17 nm$ and $\Gamma=0.055$. The reference temperature 
$\T_{ref}$ is twice the critical temperature, and, in addition, 
$k_c$ should be set to zero when the term in the bracket of equation \eqref{eq:xi} is negative.
\begin{table}
\centering
\begin{tabular}{c|ccccc}
$i$ & $N_i$        & $t_i$  & $d_i$ & $l_i$ & $\gamma_i$\\ \hline
$1$ & $1.511$      &        &       &       &     \\ 
$2$ & $2.117$      & $-1.0$ &       &       &     \\ 
$3$ & $-3.332$     & $-0.7$ &       &       &     \\ 
$4$ & $8.862$      & $0.0$  & $1$   & $0$   & $0$ \\ 
$5$ & $31.11$      & $0.03$ & $2$   & $0$   & $0$ \\  
$6$ & $-73.13$     & $0.2$  & $3$   & $1$   & $1$ \\  
$7$ & $20.03$      & $0.8$  & $4$   & $2$   & $1$ \\  
$8$ & $-0.7096$    & $0.6$  & $8$   & $2$   & $1$ \\  
$9$ & $0.2672$     & $1.9$  & $10$  & $2$   & $1$ \\  
\end{tabular}
\caption{\label{tab:coeff_ther} Coefficients and exponents of the residual thermal conductivity equation.}
\end{table}

\section{Thermal form of energy equation for a general EOS}
\label{app:II}
For the reader's convenience, the calculations needed to obtain the temperature equation for a generic equation of state, eq.~\eqref{eq:ener_evol_V}, are here explicitly reported.
The differential of the internal energy as a function of  temperature $\T^*$ and density
$\R^*$, $u^*=u(\T^*,\R^*)$, is,
\begin{align}
d u^* =\left.{\pd{u^*}{\T^*}}\right|_{\R^*} d\T^*+\left. \pd{u^*}{\R^*}\right|_{\T^*} d\R^*=
c_v^* d\T^*+\left. \pd{u^*}{\R^*}\right|_{\T^*} d\R^* \, ,
\label{eq:diff}
\end{align}
with $c_v^*=\left.{\partial u^*}/{\partial \T^*}\right|_{v^*}=\left.{\partial u^*}/{\partial \T^*}\right|_{\R^*}$ and
$v^*=1/\R^*$ the specific volume. Since $d \R^* = \R^{*2} d v^*$, it follows 
\begin{align}
\left. \pd{u^*}{\R^*}\right|_{\T^*} d \R^*= -\frac{1}{\R^{*2}}\left.\pd{u^*}{v^*}\right|_{\T^*} 
d\R^*\, .
\label{eq:dudt}
\end{align}
Combining equations \eqref{eq:dudt} and \eqref{eq:diff},  the material derivative
of the internal energy follows as
\begin{align}
\dmd{u^*}=c_v^*\dmd{\T^*}-\frac{1}{\R^{*2}}\left.\pd{u^*}{v^*}\right|_{\T^*} \dmd{\R^*}\, ,
\label{eq:total_d_I}
\end{align}
which, using the continuity equation,
${1}/{\R^*}{D\R^*/Dt}=-\ddiv \vec{u^*}$, becomes
\begin{align}
\dmd{u^*}=c_v^*\dmd{\T^*}+\frac{1}{\R^*}\left.\pd{u^*}{v^*}\right|_{\T^*} \ddiv \vec{u}^* 
=c_v^*\dmd{\T^*}+v^*\left.\pd{u^*}{v^*}\right|_{\T^*} \ddiv \vec{u^*}\, .
\label{eq:total_d_II}
\end{align}
Inserting eq.~\eqref{eq:total_d_II} in the internal energy equation~\eqref{eq:int_ener_dim}
leads to
\begin{align}
\R^* c_v^* \dmd{\T^*}=-\left(p^* + \left.\pd{u^*}{v^*}\right|_{\T^*} \right)\,\ddiv \vec{u^*}
 +\vec{\Sigma^*}:\dgrad\vec{u}^* + \ddiv\left(k^* \dgrad\T^*\right)\, .
\label{eq:ener_evol_III}
\end{align}
The first term on the right hand side of this equation can be rearranged starting from the fundamental thermodynamic relation
$ d u^* = \T^* d s^* - p^* d v^*$,
where the specific entropy can be  expressed as a function of temperature and 
specific volume $s^*=s(\T^*,v^*)$,
\begin{align}
d u^* &= \T^* \left(\left.\pd{s^*}{\T^*}\right|_{v^*} d\T^*+
\left.\pd{s^*}{v^*}\right|_{\T^*} d v^* \right) - p^* d v^*
\\
&=\left(-p^*+\T^*\,\left.\pd{s^*}{v^*}\right|_{\T^*}\right) d v^* 
 + \T^*\,\left.\pd{s^*}{\T^*}\right|_{v^*} d\T^* \ .
\label{eq:IIIterm}
\end{align}
Since the specific entropy $-s^*$ is the first derivative of the Helmholtz
specific free energy $f^*$ with respect to the temperature, one of the Maxwell's relations yieds
\begin{align}
\left.\pd{s^*}{v^*}\right|_{\T^*}=-\frac{\partial^2 f^*}{\left.\partial \T^*\right|_{v^*}
\left.\partial v^* \right|_{\T^*}}=\left.\pd{p^*}{\T^*}\right|_{v^*}
\label{eq:termI}
\end{align}
where $p^*=-\left.{\partial f^*}/{\partial v^*}\right|_{\T^*}$.
Merging equations \eqr{IIIterm} and \eqr{termI}, brings to the general identity
\begin{align}
\left.\pd{u^*}{v^*}\right|_{\T^*}=-p^*+\T^*\left.\pd{s^*}{v^*}\right|_{\T^*}=
-p^* + \T^* \left.\pd{p^*}{\T^*}\right|_{v^*}\,.
\end{align}
Inserting the above expression in the specific energy equation \eqr{ener_evol_III},
yields the required equation for the temperature field,
\begin{align}
\R^* c_v^* \dmd{\T^*}=-\T^* \left.\pd{p^*}{\T^*}\right|_{v^*} \,\ddiv \vec{u}^*
 +\vec{\Sigma^*}:\dgrad\vec{u}^* + \ddiv\left(k^* \dgrad\T^*\right)
\label{eq:ener_evol_IV}
\end{align}
which is indeed the dimensional counterpart of the dimensionless equation reported in the main text as 
eq.~\eqref{energia_n_s}.

\providecommand{\noopsort}[1]{}\providecommand{\singleletter}[1]{#1}%

\end{document}